\documentclass{aa}  
\usepackage{natbib}
\usepackage{graphicx}
\usepackage{txfonts}
\usepackage[colorlinks=true,allcolors=cyan]{hyperref}

\usepackage[normalem]{ulem}
\usepackage{xcolor}
\definecolor{ML}{rgb}{0.09,0.11,0.50} 

\begin{document} 
  \title{Properties of gas phases around cosmic filaments at $z=0$ in the IllustrisTNG simulation}
  
  \titlerunning{Gas phases in filaments}
  
  \author{Daniela Gal\'arraga-Espinosa\inst{}
    \and Nabila Aghanim\inst{}
    \and Mathieu Langer\inst{}
    \and Hideki Tanimura\inst{}
        }

  \institute{Universit\'e Paris-Saclay, CNRS, Institut d'astrophysique spatiale, 91405, Orsay, France\\
 \email{daniela.galarraga@universite-paris-saclay.fr}
        }

  \date{Received XXX; accepted YYY}

  \abstract{
   {We present the study of gas phases around cosmic-web filaments detected in the TNG300-1 hydro-dynamical simulation at redshift $z=0$. We separate the gas in five different phases according to temperature and density. 
   We show that filaments are essentially dominated by gas in the warm-hot intergalactic medium (WHIM), which accounts for more than $80 \%$ of the baryon budget at $r \sim 1$ Mpc. Apart from WHIM gas, cores of filaments ($r \leq 1$ Mpc) also host large contributions other hotter and denser gas phases, whose fractions depend on the filament population. 
   By building temperature and pressure profiles, we find that gas in filaments is isothermal up to $r \sim 1.5$ Mpc, with average temperatures of $T_\mathrm{core} = 4 - 13 \times 10^{5}$ K, depending on the large scale environment. Pressure at cores of filaments is on average $P_\mathrm{core} = 4 - 12 \times 10^{-7}$ $\mathrm{keV.cm}^{-3}$, which is $\sim 1000$ times lower than pressure measured in observed clusters. We also estimate that the observed Sunyaev-Zel'dovich (SZ) signal from cores of filaments should range between $0.5 < y < 4.1 \times 10^{-8}$, and these results are compared with recent observations.
   Our findings show that the state of the gas in filaments depend on the presence of haloes, and on the large scale environment.}}


\maketitle



\section{Introduction}

Under the action of gravity, matter is assembled to form a gigantic network composed of nodes, filaments, walls and voids called the cosmic web \citep{Lapparent1986, Bond1996}. This structure is mainly set by the dynamics of dark matter (DM), which builds the cosmic skeleton, and baryonic matter is accreted onto the dark matter skeleton driven by gravity. Studies based on simulations have shown that cosmic filaments might contain almost $50\%$ of the mass of the Universe \citep[e.g.][]{Cautin2014}, so our understanding of matter in the Universe is closely tied to that of filaments.

Besides gravity and due to their collisional nature, baryons are also subject to a large variety of physical processes such as heating, cooling, mass ejection and shocks caused, for example, by stellar and AGN feedback, galactic winds, radiation, etc . These processes form a complex cycle in which gas is pushed into different physical states, or phases, that co-exist in the cosmic web and that are spatially correlated with the different structures at different cosmic scales \citep[e.g.][]{Ursino2010, Shull2012, CenOstriker2006, Haider2016ILLUSTRIS, Martizzi2019a, Ramsoy2021}.

An ideal tool to explore gas in filaments without the difficulties specific to observational data analysis (e.g. foreground contamination, low signal to noise ratio, redshift uncertainties, etc.) are hydro-dynamical numerical simulations. Indeed, these have significantly helped to better understand the properties of baryons and the processes to which they are subjected in the cosmic web \citep[e.g.][]{Ursino2010, Shull2012, Nevalainen2015, GhellerVazza2015, GhellerVazza2016}. 
For example, \cite{GhellerVazza2019_surveyTandNTprops_fils} investigated the link between the properties of cosmic filaments (e.g. mass, length, temperature, etc.) and the physics deciding the fate of baryons in the cosmic web. \cite{Martizzi2019a} analysed the gas content and separated gas into different phases of the various structures of the cosmic web at different redshifts of the IllustrisTNG simulation \citep{Nelson2019}. More recently, \cite{GhellerVazza2020} have shown that one of the most promising tracers of gas in the diffuse warm-hot intergalactic medium (WHIM) around filaments is the SZ effect, thanks to its linear dependence on gas pressure.

Given that cosmic filaments lack a standard and unique definition, and that their mass densities are low in comparison with the nodes of the cosmic web, the studies of filaments face all the same limitations in terms of detection and identification. Several algorithms have been developed in order to overcome these limitations, and they detect filaments using different approaches. For example, filaments can be defined based on the topology of the matter density field, on density thresholds, on the velocity field, or even with a probabilistic approach \citep[see for example][]{AragonCalvo2010spineweb, Disperse_paper1, Sousbie2011b,Cautun2013nexus, Libeskind2018, Bonnaire2020Trex, Pereyra2020_semita}.

In the present paper, we study gas around cosmic filaments detected with the DisPerSE algorithm \citep{Disperse_paper1, Sousbie2011b} in the galaxy distribution of the IllustrisTNG simulation \citep{Nelson2019} at redshift zero. We separate the gas in different density and temperature states, called phases, and we study the gas properties around the different populations of filaments identified in \cite{GalarragaEspinosa2020}. We map the distribution of the different baryonic phases around filaments by building phase fraction profiles, and we characterise their thermodynamical properties by means of temperature and pressure profiles. We remove the contamination of gas contained in nodes, and we identify the contribution of massive ($M_\mathrm{tot} > 10^{12} \, \mathrm{M}_\odot$) galactic haloes in order to explicitly focus on the properties of the gas of filaments that is not associated with smaller-scales collapsed structures.

In Section \ref{Sect:Data}, we introduce the IllustrisTNG simulation and we describe the filament catalogue as well as the five different phases studied in this work. We show our results on the distribution of these phases around the different types of filaments in Sect.~\ref{Sect:results1}. Section~\ref{Sect:Results2} presents a study of the temperature and pressure of the five gas phases around filaments. We analyse, in Sect.~\ref{Sect:ContributionGalaxies}, the contribution of gas in galactic haloes to the filament temperature and pressure when the haloes are not removed in the analysis. Finally, we estimate the SZ signal from filaments in Sect.~\ref{Sect:SZ}, and we summarise our conclusions in Sect.~\ref{Sect:Conclusions}.

\section{\label{Sect:Data}Data}

\subsection{\label{SubSect:TNGsimu_description}The IllustrisTNG simulation}

We analyse the gravo-magnetohydrodynamical simulation IllustrisTNG\footnote{https://www.tng-project.org} \citep{Nelson2019}. This simulation follows the coupled evolution of DM, gas, stars, and black holes from redshift $z=127$ to $z=0$. IllustrisTNG is run with the moving-mesh code Arepo \citep{Arepo} with the cosmological parameters from \citep{Planck2015Cosmo}, namely $\Omega_{\Lambda,0} = 0.6911$, $\Omega_{m,0}=0.3089$, $\Omega_{b,0}=0.0486$, $\sigma_{8}=0.8159$, $n_s=0.9667$ and $h=0.6774$.

In this work, we choose to use the largest simulation volume with the highest mass resolution of the IllustrisTNG suite in order to accurately describe large cosmic filaments and their gas content down to small scales. We therefore focus on the simulation box TNG300-1 at a redshift $z=0$. It consists in a cube of around 302 Mpc side length with a DM resolution of $m_{\mathrm{DM}} = 4.0 \times 10^{7} \mathrm{M_{\odot}}/h$ and a number of particles of $N_{\mathrm{DM}} =2500^{3}$.

The gaseous component in the IllustrisTNG simulation is implemented by Voronoi cells evolving in time using Godunov's method \citep{Nelson2019}. Thanks to Riemann solvers at the cell interfaces, the Voronoi cells track the conserved quantities of the gas fluid (e.g. mass, momentum, energy), and they are refined and de-refined according to a mass target of $7.6 \times 10^6 \, \mathrm{M}_\odot /h$ \citep{Arepo, Nelson2019, Weinberger2020arepo, Pillepich2018TNGmodel}. Finally, the baryonic processes driving gas dynamics (like star formation, stellar evolution, chemical enrichment, gas cooling, black hole formation, growth, and feedback, etc.) are implemented in a subgrid manner following the `TNG model' \citep{Pillepich2018TNGmodel, Nelson2019}. This model was specifically calibrated on observational data to match to the observed galaxy properties and statistics \citep{Nelson2019}. Its complete description can be found in \cite{Pillepich2018TNGmodel}.

\subsection{\label{SubSect:fil_cata}Filament catalogue}

The catalogue of filaments used in this work was constructed in \cite{GalarragaEspinosa2020} to which we refer the reader for details and technicalities. Here, we simply summarise the main important points.

We detect the filamentary structures in the simulation box by applying DisPerSE \citep{Disperse_paper1, Sousbie2011b} to the galaxy catalogue of the TNG300-1 simulation at $z = 0$. Galaxies of this catalogue have been selected to be in the stellar masses range of $10^{9} \le \mathrm{M}_{*} [\mathrm{M_{\odot}}] \le 10^{12}$, in order to correspond to observational limits \citep{Brinchmann2004, Taylor2011}.
From the galaxy distribution, DiPerSE first computes the density field using the Delaunay Tessellation Field Estimator \citep[DTFE,][]{SchaapWeygaert2000, WeygaertSchaap2009}.
In order to minimise the contamination by shot noise, and to prevent the identification of small scale, possibly spurious features, following \cite{Malavasi2020_sdss}, we smooth the density field by averaging the density value at each vertex of the Delaunay tessellation with the values at the surrounding vertices. DisPerSE then identifies the critical points of the density field (points where the gradient of the field is zero) using Discrete Morse theory. Filaments are thus defined as sets of segments connecting maximum density critical points to saddles. The user can choose the significance of the detected filaments by fixing the persistence ratio of the corresponding pair of critical points and, following \cite{GalarragaEspinosa2020}, the persistence is fixed to $3\sigma$ in this work.

The filaments of the obtained catalogue have a maximum length of $L_f = 65.6$ Mpc, their minimum is $L_f = 0.4$ Mpc, and the mean and median lengths are respectively 10.9 and 8.8 Mpc.
As in \cite{GalarragaEspinosa2020}, we identify two statistically distinct populations of filaments, short ($L_f < 9$ Mpc) and long ($L_f \geq 20$ Mpc), depending on the density of their environments in the cosmic web (respectively denser and less-dense). In addition, we consider medium-length filaments ($9 \leq L_f < 20$ Mpc) in order to study gas around all the filaments of our catalogue. 
These specific length boundaries are not sharply defined, as they weekly depend on the density of the tracers of the matter distribution, that is in our case the number of galaxies in the simulated volume ($10^{-2}$ gal/$\mathrm{Mpc}^3$). However, the results we show here no dot depend on the exact values of these length boundaries.

\subsection{\label{SubSect:gas_catalogue_building}Gas cell dataset}

\begin{figure}
    \centering
   \includegraphics[width=0.5\textwidth]{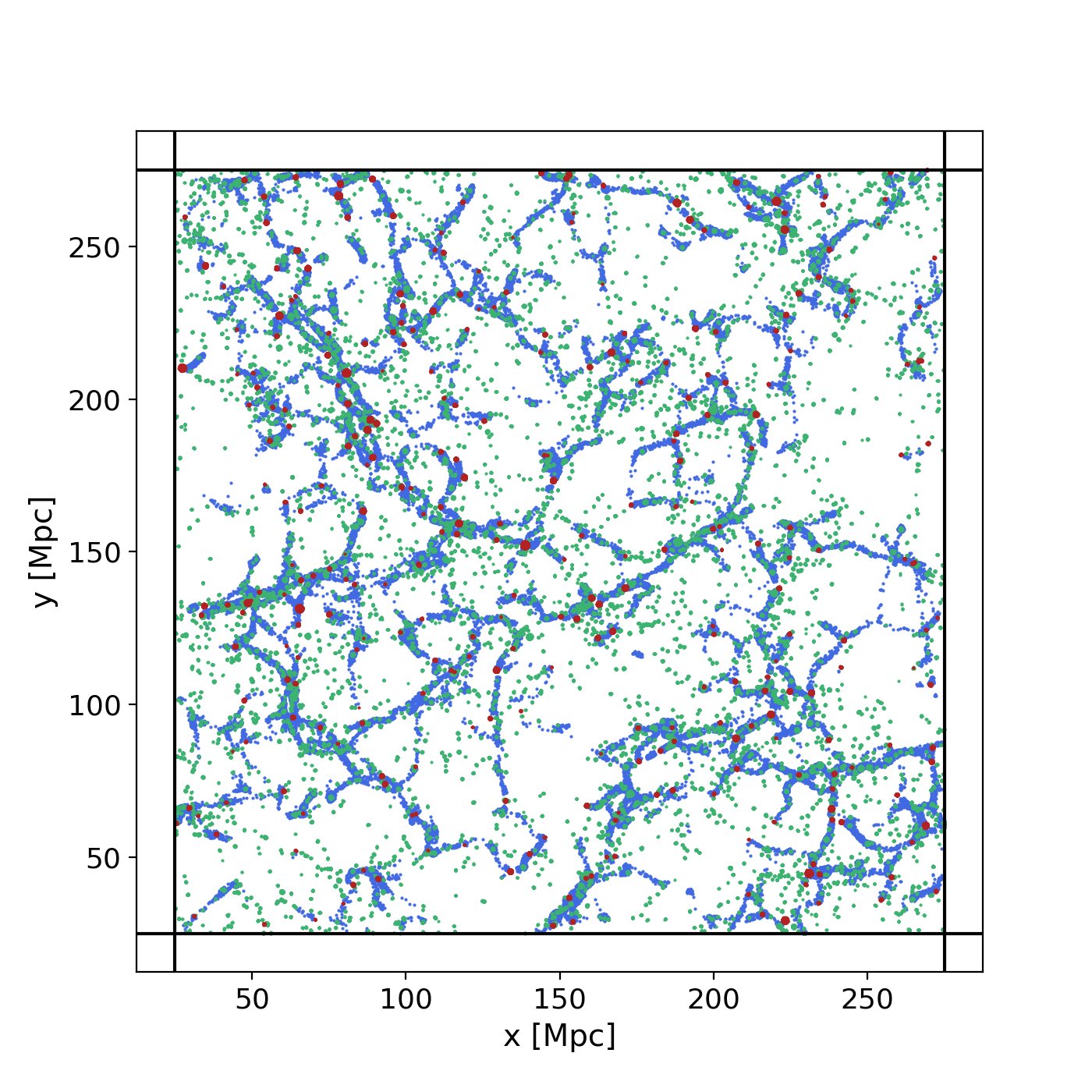}
   \caption{2D projection of a slice of thickness 50 Mpc in the $xy$ plane of the TNG300-1 box. The red points correspond to gas cells within the $R_{200}$ radii of nodes (traced in this work by the DisPerSE maximum density critical points, see Sect.~\ref{SubSect:gas_catalogue_building}). The blue points represent gas around filaments, at distances closer than 1 Mpc from the spine of the cosmic web. Finally, the green points show gas in the $R_{200}$ region of galactic haloes (i.e. within spheres of radius $1 \times R_{200}$). The galactic haloes residing in filaments are clearly apparent (green over blue regions). Note that in this figure the centres of the cells are plotted as points, but one should keep in mind that gas in the TNG300-1 simulation is described by Voronoi cells having various volumes.}
    \label{Fig:xy_2D}
    \end{figure}

The TNG300-1 simulation box provides a dataset of $15,625,000,000$ gas cells. As we want to study gas properties that are associated with filaments, we remove from the analysis the cells located within spheres of radius $3 \times R_{200}$ centered on the position of the maximum density critical points of the Delaunay density field (hereafter CPmax, see Sect.~\ref{SubSect:fil_cata}), that are the topological nodes of the skeleton. This step removes $11\%$ of the initial dataset. For illustration, the gas cells within $1 \times R_{200}$ of the CPmax are represented by the red points in the $xy$ slice of Fig.~\ref{Fig:xy_2D}.

We also remove the contribution of galactic haloes, that we define as massive ($M_\mathrm{tot} > 10^{12} \, \mathrm{M}_\odot$) collapsed structures hosting at least one galaxy of stellar mass $10^{9} \le \mathrm{M}_{*} \le 10^{12} \, \mathrm{M_{\odot}}$ (see green points in Fig.~\ref{Fig:xy_2D}). We therefore discard the gas cells located within spheres of radius $3 \times R_{200}$ centred on these structures. After these steps, our dataset contains $76\%$ of the initial cells, the other $24 \%$ lying within $3 \times R_{200}$ of nodes ($11 \%$) or galactic haloes ($13 \%$).

Furthermore, given the excess of critical points near the boundaries of the simulation box, we chose, following \cite{GalarragaEspinosa2020}, to completely disregard the regions of thickness 25 Mpc at the box boundaries (vertical and horizontal lines in Fig.~\ref{Fig:xy_2D}), and to replicate the cell distribution at the borders, in order to reduce the effects of a volume-limited simulation in the gas profiles.

By masking the CPmax and the galactic haloes, and by removing the cells at the borders, we obtain a dataset containing $\sim 6,400,000,000$ gas cells (i.e. $\sim 40 \%$ of the initial dataset). This is still an extremely large number, and dealing with such an amount of information is computationally very expensive and time consuming. So, in order to perform our analysis in reasonable computing time, we sample the gas dataset by randomly selecting one cell out of 500. By doing so, we end up with a final set containing $12,711,272$ cells.
We have checked that this number is representative of the full cell dataset, as different random samples give exactly the same results in terms of gas distributions and statistical properties. Hence, the resulting gas dataset allows us to perform the analysis of properties of gas of the inter-filamentary medium, without the contamination of gas in haloes and nodes. This gas dataset will be used in the following Sects.~\ref{Sect:results1} and \ref{Sect:Results2}, while in Sect.~\ref{Sect:ContributionGalaxies}, gas in haloes will be included to show explicitly how the latter contribute to the baryon content, temperature and pressure profiles of filaments.

\subsection{\label{SubSect:cell_TandP_method}Cell temperature and pressure}

The IllustrisTNG simulation offers information about various properties and characteristics of the gas cells (labelled as \texttt{PartType0}). We derive the temperature and pressure of each cell by employing their corresponding \texttt{InternalEnergy}, \texttt{ElectronAbundance} and \texttt{Density} fields, hereafter $u$, $x_e$ and $\rho_\mathrm{cell}$ respectively.

Gas cell temperature is computed under the assumption of perfect monoatomic gas, as:

\begin{equation}
    T_\mathrm{cell} = (\gamma -1) \times \frac{u}{k_\mathrm{B}} \times \mu.
\end{equation}
Here $\gamma$ corresponds to the adiabatic index of perfect monoatomic gas ($\gamma = 5/3$), $k_\mathrm{B}$ is the Boltzmann constant, and $\mu$ denotes the mean molecular weight. The latter is estimated as follows:
\begin{equation}
    \mu = \frac{4}{1 + 3 X_\mathrm{H} + 4 X_\mathrm{H} x_e} \times m_\mathrm{p},
\end{equation}
where $X_\mathrm{H}$ is the hydrogen mass fraction ($X_\mathrm{H} = 0.76$), and $m_\mathrm{p}$ denotes the mass of the proton. In these equations, all the quantities are expressed in the International System of Units.\\

The thermodynamic pressure is estimated from the temperature and the electron density, as:
\begin{equation}
    P_\mathrm{cell} = n_e \, k_\mathrm{B} \, T_\mathrm{cell}.
\end{equation}
The electron density $n_e$, in $\mathrm{m}^{-3}$, is computed as the 
electron abundance times the hydrogen density, $n_e = x_e \times n_\mathrm{H}$, where $n_\mathrm{H} = X_\mathrm{H} \,\, \rho_\mathrm{cell}/m_\mathrm{p}$. We convert the pressure from pascals to the more usual unit in astrophysics of $\mathrm{kev}.\mathrm{cm}^{-3}$.

\subsection{\label{SubSect:gas_phases_definition}Definition of gas phases}

Gas in the cosmic web is subjected to several complex and interconnected physical processes, that can be essentially summarised by accretion, ejection, heating and cooling. These processes are not isotropic, they depend on the local environment in the cosmic web, and so they  distribute the gas into different phases. Phases can easily be identified with the help of a phase-space diagram showing the relation between gas temperature and gas density  \citep[e.g.][]{Ursino2010, Shull2012, CenOstriker2006, Haider2016ILLUSTRIS, Martizzi2019a}. These phases share neither the same physical properties, nor the same spatial distribution as we will show below.

\begin{figure}
    \centering
   \includegraphics[width=0.5\textwidth]{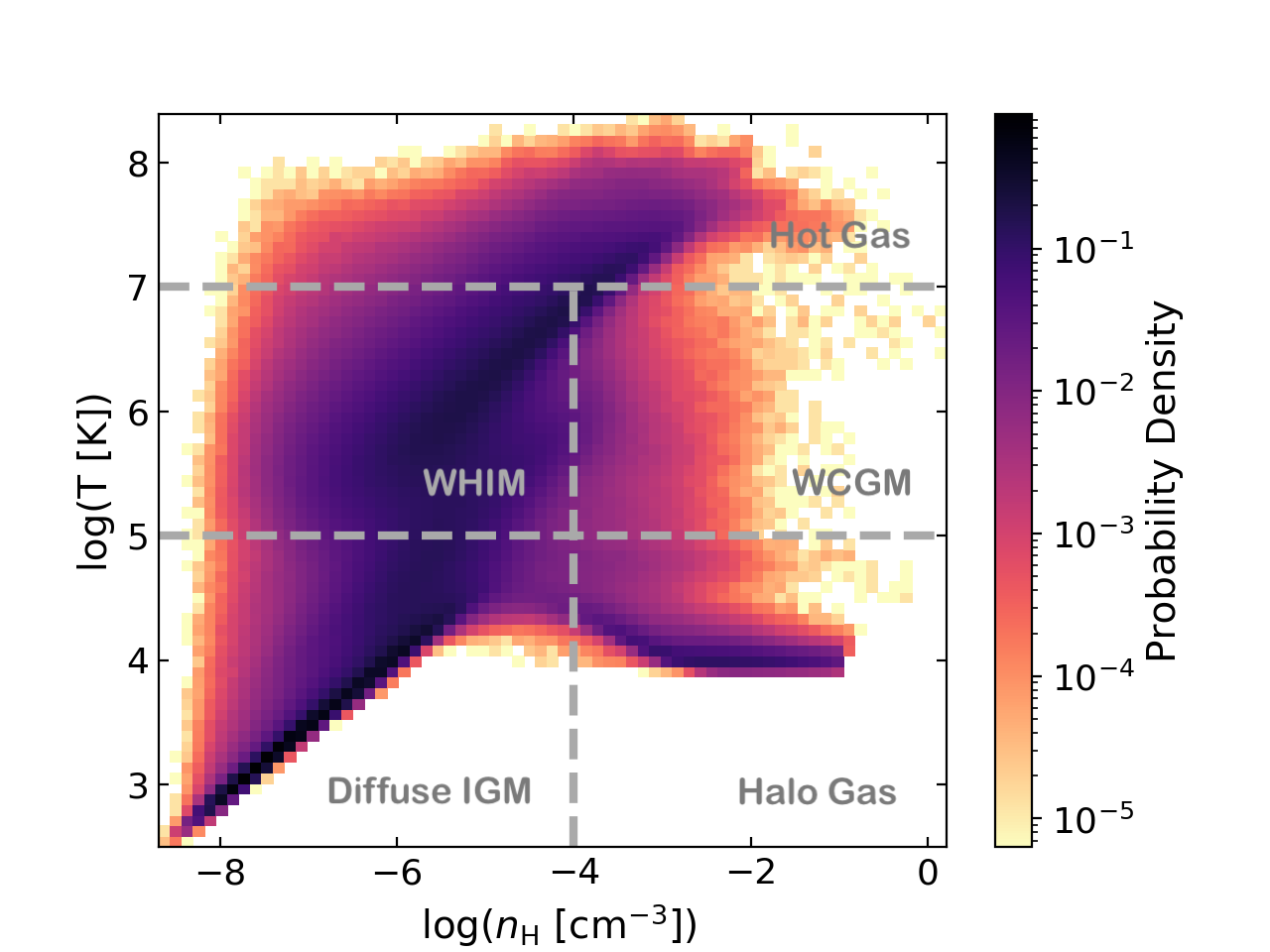}
   \caption{Phase-space of the gas cells in the TNG300-1 simulation, shown by the normalised 2D histogram of the cells. The five different gas phases studied in this work are delimited by the thick dashed lines. They are defined as in \cite{Martizzi2019a}, and presented in Table~\ref{Table:gas_phases} and Sect.~\ref{SubSect:gas_phases_definition}}.
    \label{Fig:gas_phase_definition}
    \end{figure}

In this work, we analyse the baryonic gas not only in terms of the various populations of cosmic filaments we identified, but also in terms of its distribution across five different phases. Indeed, we consider the five phases of gas presented in Table~\ref{Table:gas_phases} and in Fig.~\ref{Fig:gas_phase_definition}, that shows the normalised 2D histogram of the gas cells of the TNG300-1 simulation. The five gas phases are defined according to their hydrogen number density\footnote{We note that the hydrogen number density is a direct tracer of the total gas mass density $\rho_\mathrm{cell}$, as $n_\mathrm{H} = X_\mathrm{H} \,\, \rho_\mathrm{cell}/m_\mathrm{p}$.}, $n_\mathrm{H}$, and their temperature. These are the diffuse intergalactic medium (Diffuse IGM), the warm-hot intergalactic medium (WHIM), the warm circumgalactic medium (WCGM), the halo gas and the hot gas. For further details, we refer the reader to Sect. 2.3 of \cite{Martizzi2019a}.
We note, however, that in our analysis we do not consider star-forming gas, that is defined in \cite{Martizzi2019a} as gas with densities higher than $n_\mathrm{H} > 0.13$ $\mathrm{cm}^{-3}$, temperatures lower than $T < 10^{7}$ K and star formation rate SFR > 0. 
This is motivated by the fact that the \texttt{ElectronAbundance} field (see Sect.~\ref{SubSect:cell_TandP_method}) of star-forming cells is altered by the subgrid model of star-formation \citep{SpringelHernquist2003_SFRmodel} employed in the simulation \cite[see an example in][]{Pakmor2018_FaradayRotationTNG}. Moreover, it is worth noticing that star formation is not a representative property of gas in the IllustrisTNG simulation, as cells with SFR > 0 account for only a tiny fraction of the total gas budget ($\sim 0.1 \%$).

\begin{table*}
\caption{Definition of the five different gas phases studied here (see Sect.~\ref{SubSect:gas_phases_definition} and Fig.~\ref{Fig:gas_phase_definition}). Further details can be found in \cite{Martizzi2019a}.}
\label{Table:gas_phases}     
\centering  
\begin{tabular}{ c | c c | l }
 \hline\hline  
     & Density [$\mathrm{cm}^{-3}$] & Temperature [K] & Comments \\ \hline
    Diffuse IGM & $n_\mathrm{H} \leq 10^{-4}$ & $T \leq 10^{5}$ & Gas in the lowest-density regions of the cosmic web.
\\
    WHIM & $n_\mathrm{H} \leq 10^{-4}$ & $10^{5} < T \leq 10^{7}$ & Gas that has been accreted onto cosmic structures and heated by shocks.
\\
    WCGM & $n_\mathrm{H} > 10^{-4}$ & $10^{5} < T \leq 10^{7}$ & In the surroundings of galaxies, sensitive to galactic physics.
\\
    Halo Gas & $n_\mathrm{H} > 10^{-4}$ & $T \leq 10^{5}$ & In the interstellar medium of galaxies, located inside or near them.
\\
    Hot Gas & no cut & $T > 10^{7}$ & Shock-heated gas located in the denser regions of the cosmic web.
\\
  \hline
 \end{tabular}
\end{table*}

Before presenting our results in the next sections, we would like to recall that the boundaries of these phases are somewhat artificial, and should not be considered as sharp limits. Gas cells in the neighbourhood of these limits may suddenly move from one phase to the other under the influence of even the tiniest perturbation (for example, WCGM gas just needs to be heated up to temperatures beyond $10^7$ K to be counted as hot gas). Similarly, gas cells at a given place in the phase diagram may have reached their location through a large variety of processes. For instance, gas within the boundaries of the WCGM phase may correspond either to WHIM gas that has been accreted and has become denser, to halo gas that has been heated by feedback effects, or to hot gas that has cooled down.
Therefore these density and temperature limits defining the gas phases should simply be considered as guiding references.  


\section{\label{Sect:results1}Distribution of gas phases around filaments}

\subsection{\label{SubSect:gas_fraction_profiles}Gas fraction profiles}

We describe here how the different gas phases introduced above are distributed around the filaments of our catalogue (Sect.~\ref{SubSect:fil_cata}). We compute radial profiles of the mass fraction of each gas phase (hereafter called $\varphi_i$) along the direction perpendicular to the filament spine (hereafter called $r$). The phase mass fraction is defined as
\begin{equation}\label{Eq:phi_i}
    \varphi_i(r) = \frac{\rho_{\mathrm{gas},i} (r)}{\rho_{\mathrm{gas}, \mathrm{tot}}(r)},
\end{equation}
where $\rho_{\mathrm{gas},i}(r)$ corresponds to the mass density of the gas phase $i$ ($i =$ WHIM, WCGM, etc.) at a given location around filaments, and $\rho_{\mathrm{gas}, \mathrm{tot}}(r)$ is the total mass density of gas (all phases included) at this location. 
These densities are computed as mass averages in volumes of concentric cylindrical shells (denoted by the index $k$) around the axis of filaments. For all the $N_\mathrm{seg}$ filament segments, for a given distance $r_k$ from their axes, we sum the masses of the gas cells $m_{\mathrm{cell}}$ located inside the cylindrical shells of outer radius $r_{k}$ and of thickness $r_{k} - r_{k-1}$, in order to get the total enclosed gas mass at this location. We then divide this result by the corresponding volume, that is the volume of the hollow cylinder of thickness $r_{k} - r_{k-1}$ and of height the total segment length. This is shown by:

\begin{equation}\label{Eq:RHOprofile}
    \rho_{\mathrm{gas}}(r_k) = \frac{\displaystyle
    \sum_{s = 1}^{N_\mathrm{seg}}
   \left( \sum_{j = 1}^{N(k)} m_{\mathrm{cell}, j} \right)_{s}
    }{
    \pi (r^2_{k} - r^2_{k-1}) \,
    \displaystyle \sum_{s = 1}^{N_\mathrm{seg}} l_s 
    } \, ,
\end{equation}
where, $N(k)$ corresponds to the number of cells within the $k$-th shell around the segment $s$ of length $l_s$. We note that the radial distance $r$ is binned in $k = 20$ equally spaced logarithmic bins, starting from the filament spine up to $100$ Mpc. \\

    \begin{figure*}
   \centering
   \includegraphics[width=1\textwidth]{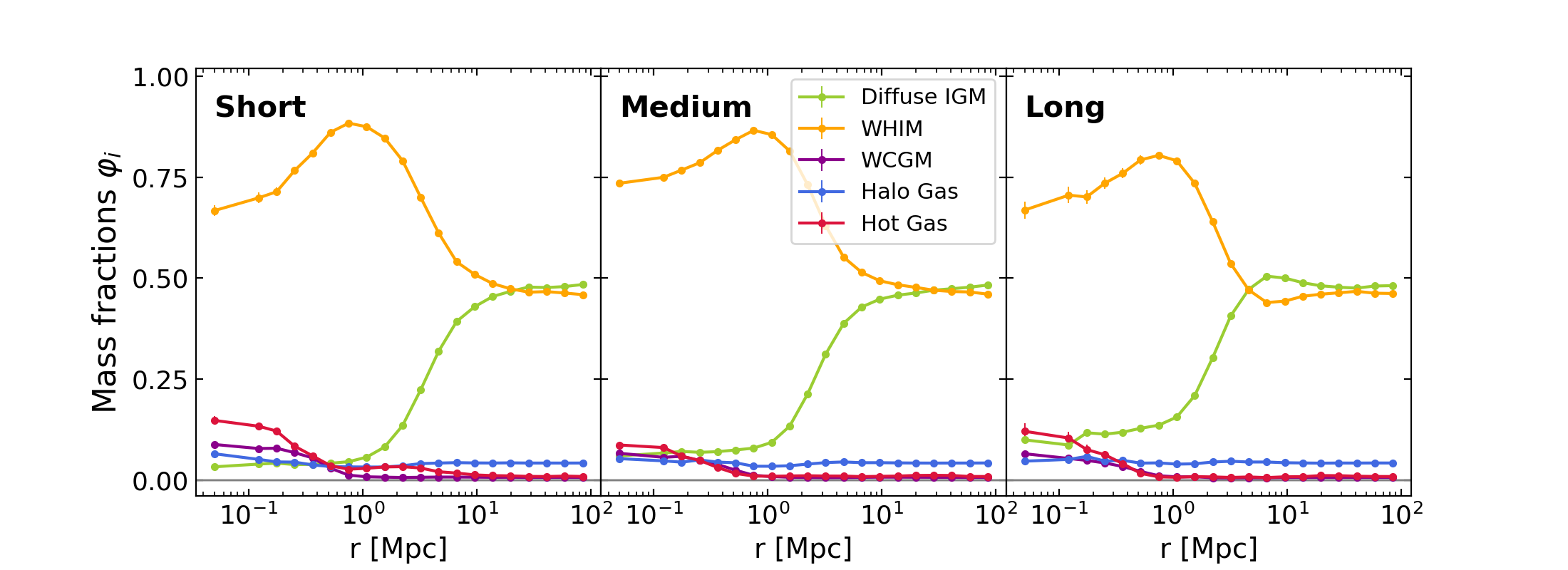}
   \caption{Gas mass fraction profiles $\varphi_i$ (see Eq.~\ref{Eq:phi_i}) of the five different phases (labelled by the index $i$) around short (left panel), medium-length (centre) and long filaments (right panel). All the gas phases used in this work are presented in Sect.~\ref{SubSect:gas_phases_definition}.}
    \label{Fig:gas_phase_fraction}
    \end{figure*}

The resulting phase fraction profiles $\varphi_i$ are presented in Fig.~\ref{Fig:gas_phase_fraction} for short, medium-length and long filaments, respectively in the left, middle and right panels. The error bars in this figure are obtained by propagating these of the density measurements. The latter are obtained using the bootstrap method, i.e. we compute the standard deviation of $1000$ density profiles, obtained by applying Eq.~\ref{Eq:RHOprofile} to a set of $N_\mathrm{seg} = 1000$ randomly extracted (with replacement) filament segments. These errors thus quantify the statistical variance of the density around our limited sample of filaments.
In this figure, we distinguish three different regimes for all the filaments:

\textit{(i)} Far away from the filament spine (at distances $r>10$ Mpc), gas is mainly in a diffuse state (more than $95 \%$ of the total budget). This is shown by the fractions of Diffuse IGM and WHIM gas (respectively in green and yellow), that are the highest in all filaments and, as expected, Diffuse IGM is the most representative phase ($\sim 50 \%$) at such large distances.
In the case of short filaments (left panel of Fig.~\ref{Fig:gas_phase_fraction}), from outside to inside, we see a decline of the Diffuse IGM fraction starting at distances of $\sim 20$ Mpc from the spine, and this goes along with a rise of the fraction of WHIM. This feature can be interpreted as the accretion of diffuse gas to filaments, that start at these large distances in the short population. Accreted gas is gravitationally heated (as we will see in the temperature profiles of Fig.~\ref{Fig:Temp_pro}), and Diffuse IGM gas is thus converted to WHIM, which becomes the dominant phase at $r \sim 20$ Mpc from the spine of the filament (see intersection point).
However, in long filaments (right panel), this accretion feature happens at distances much closer to the spine, as the Diffuse IGM and WHIM fractions intersect only at $r \sim 5$ Mpc.
These differences are likely due to the fact that short and long filaments reside in different environments of the cosmic web \citep{GalarragaEspinosa2020}. Short filaments are in denser regions (e.g. in the vicinity of galaxy clusters), where the heating of the gas due to accretion happens at larger radii from the spine than in less dense environments (e.g. in the surroundings of voids), which are traced by long filaments.
Finally, the fractions of WCGM, Halo and Hot gas are only tiny at these large distances from the spines.

\textit{(ii)} At intermediate distances ($1<r \leq 10$ Mpc), the accretion of cold and diffuse gas to the core continues. We observe a very significant decrease of the fraction of Diffuse IGM, along with an abrupt increase of that of the WHIM phase. 
The fraction of WHIM reaches its maximum ($\sim 88 \%$ of the total baryon budget in short, and $\sim 80 \%$ in long filaments) at distances of $r \sim 1$ Mpc from the spine. This radial scale of 1 Mpc is independent from the length of the filament, and does not depend on the resolution of the simulation either (see Appendix~\ref{Appendix:Resolution} for details), thus pointing out to a more general radial extent of gas in filaments, that might be defined as the radius where baryonic processes in filaments (e.g. shocks, feedback effects, mergers) start counteracting the pure gravitational infall of gas. Moreover, in this radial regime of $1<r \leq 10$ Mpc, almost all Diffuse IGM gas has been heated and turned into another phase, becoming negligible (less than $5 \%$ and decreasing) in the total baryon budget, which is completely dominated by WHIM.
There are no significant changes in the denser phases (WCGM, Halo and Hot Gas), whose fractions remain constant and tiny on the outskirts of filaments.

\textit{(iii)} The cores of filaments ($r \leq 1$ Mpc) are characterized by a very sharp decrease of the fraction of WHIM and the increase of the WCGM and Hot gas, that remained tiny so far. Precisely, we report in Table~\ref{Table:fractions_core} the fractions of each gas phase in the first radial bin, i.e. $r=0.05$ Mpc. We notice that, despite the decrease, WHIM remains the dominant phase in all types of filaments (accounting for $\sim 70 \%$ of the baryons), which is in qualitative agreement with previous findings at $z=0$, using different simulations and filament finders \citep[e.g.][]{Nevalainen2015, Cui2018, Cui2019, Martizzi2019a, Tuominen2020}. Note that cores of filaments are also composed of hotter and denser gas (WCGM and especially Hot gas). Indeed, as WHIM gas approaches or falls into the filament, it is subjected to a variety of baryonic processes (shocks, feedback effects, mergers, etc.), specific to these dense environments which contain significantly larger numbers of galaxies than the outskirts of filaments \citep[as shown in][]{GalarragaEspinosa2020}, thus making them susceptible to the effects of galactic physics (see Sect.~\ref{Sect:ContributionGalaxies}).
Moreover, we note that the fractions of the different phases depend on the length of filament, and this is further investigated in the next subsection, where we study, at the cell level, the gas content of cores of the short and long populations.

\begin{table}
\caption{Mass fractions $\varphi_i$ of the different gas phases at the core of short, medium-length and long filaments. The values here are given in $\%$, and correspond to the first radial bin of the $\varphi_i$ profiles of Fig.~\ref{Fig:gas_phase_fraction}, that is $r=0.05$ Mpc}.
\label{Table:fractions_core}     
\centering  
\begin{tabular}{ c | c c c }
 \hline\hline  
     & Short & Medium & Long \\ \hline
    Diffuse IGM & 
    $(3 \pm 1)\%$ & 
    $(6 \pm 1)\%$ &
    $(10 \pm 1) \%$\\
    WHIM & 
    $(67 \pm 1)\%$ & 
    $(73 \pm 1)\%$ &
    $(67 \pm 2)\%$ \\
    WCGM & 
    $(9 \pm 1)\%$ &
    $(7 \pm 1)\%$ &
    $(6 \pm 1)\%$\\
    Halo Gas & 
    $(6 \pm 1)\%$ & 
    $(5 \pm 1)\%$ & 
    $(5 \pm 1)\%$\\
    Hot Gas & 
    $(15 \pm 1)\%$ & 
    $(9 \pm 1)\%$ & 
    $(12 \pm 2)\%$\\
  \hline
 \end{tabular}
\end{table}

\subsection{\label{SubSect:Phase_Spaces}Phase diagrams of gas in filaments}

    \begin{figure*}
   \centering
   \includegraphics[width=1\textwidth]{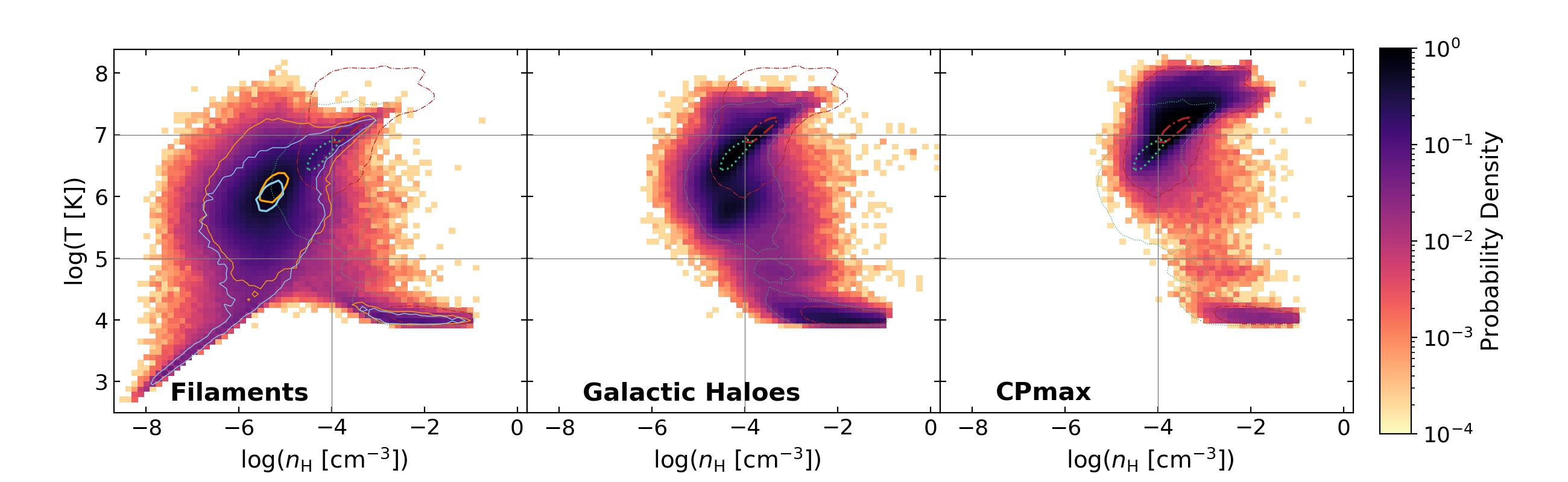}
   \caption{Phase portraits of gas, presenting the normalised 2D histograms of the gas cells in the corresponding structures. \textit{First panel}: gas in filaments (within 1 Mpc from the core). \textit{Second panel}: gas within the $R_{200}$ radius of galactic haloes (defined as haloes of mass $M_\mathrm{tot} > 10^{12} \, \mathrm{M}_\odot$ hosting at least one galaxy of mass $10^{9} \le \mathrm{M}_{*} [\mathrm{M_{\odot}}] \le 10^{12}$). \textit{Third panel}: gas within the $R_{200}$ radius of the  DisPerSE maximum density critical points (CPmax), tracers of nodes. The $68\%$ and $99\%$ contours correspond to results of short (orange) and long filaments (blue), to galactic haloes (dotted green lines), and CPmax (dashed-dotted red contours).}
    \label{Fig:PS_3cols}
    \end{figure*}
    
In order to have a clearer picture of the physical properties of gas in filaments, we now analyse their gas content by the means of phase diagrams.
Indeed, as we mentioned above, the boundaries between the five phases used in Fig.~\ref{Fig:gas_phase_fraction} are somehow arbitrary, and must not be considered as sharp limits. This is particularly the case in cores of filaments (and in haloes and nodes) due to the numerous physical processes affecting the gas in these dense regions.

We build the phase diagrams of gas in cylinders of radius 1 Mpc around the spine of filaments. This radius has been chosen following the results of Fig.~ \ref{Fig:gas_phase_fraction}, as it corresponds to the distance from which the WHIM fraction starts decreasing, giving room to the other hotter and denser baryonic phases.
The phase diagram of cells in filaments is presented in the 2D histogram of Fig.~\ref{Fig:PS_3cols} (first panel).
For the sake of comparison, we also analyse the phase diagram of gas inside galactic haloes (see definition in Sect.~\ref{SubSect:gas_catalogue_building}) and around the DisPerSE CPmax points, which are not counted in the first filament phase diagram. These are presented respectively in the second and third panels of Fig.~\ref{Fig:PS_3cols}, where the gas cells within spheres of radius $1 \times R_{200}$ centered on galactic haloes and on CPmax are displayed.

The phase portrait of gas in filaments  shows that most of the gas cells in these structures have temperatures and densities corresponding to the WHIM phase, which is in agreement with the previous results of Fig.~\ref{Fig:gas_phase_fraction} and Table~\ref{Table:fractions_core}. 
The contours of short and long filaments (orange and blue lines) are rather circular and share the common region around $n_\mathrm{H} = 10^{-5.3}$ $\mathrm{cm}^{-3}$ and $T=10^{6.1}$ K. However, their overall distribution is significantly different, showing that these two populations of filaments are not made by exactly the same gas. 
Indeed, contours of long filaments are shifted towards the cooler values with respect to short filaments, and they present a contribution of cold and diffuse IGM gas. The latter corresponds to primordial gas that has never been heated by baryonic processes, and whose temperature is regulated by the competition between radiative heating (by the UV background) and the expansion of the Universe \citep{Valageas2003}. On the other hand, short filaments present larger contributions of hot gas, as expected from these puffy and dense structures.
Let us now comment on the small, but non-negligible, presence of cold and dense gas (Halo gas), in the bottom right of the phase portrait of filaments, as it might play a major role in the star formation of galaxies in filaments. Indeed, it is known that cores of filaments are mainly populated by quiescent galaxies, that have ceased forming stars \citep[e.g.][]{Malavasi2017, Kraljic2018, Kraljic2019, Bonjean2020filaments}. However, recent studies have claimed the detection of a slight increase star formation in galaxies located in cores of filaments, at distances lower than $r<1$ Mpc \citep{LiaoGao2019, Singh2020}. According to \cite{LiaoGao2019}, this surprising burst of star formation might be explained by the presence of filaments, that can feed galaxies residing at their cores with cool and dense gas (at the galaxy outskirts). This gas might correspond to the Halo gas revealed in this work, and which we have checked that is not a contribution of massive galactic haloes (see Fig.~\ref{Fig:PS_fils_haloesincluded} in Sect.~\ref{Sect:ContributionGalaxies}). Cold and dense gas is a reservoir of star formation for galaxies in the cores of filaments \citep[see in particular][]{Singh2020}.\\

We compare gas in filaments with gas in galactic haloes and around CPmax (second and third panels of Fig.~\ref{Fig:PS_3cols}). Galactic haloes show a large amount of their cells in an elliptical pattern at temperatures from $10^{6.5}$ K to $10^{7}$ K, that forms an elongated contour from WHIM to Hot gas. As expected, there is also a significant contribution of cold and dense gas (Halo gas) in these structures. 
Gas around the CPmax, the tracers of nodes in this work (last panel) also exhibits the elliptical contour mentioned above, but the most striking feature of this phase portrait is the significant excess towards the hottest and densest parts of the plot, in the Hot gas domain. This hot gas is expected around these critical points, that coincide with the densest structures of the universe where the shock-heating processes are the most efficient.
We note that only a tiny fraction of gas in the WHIM region is detected around the CPmax, and that Diffuse IGM gas is absent near these points and as well as in galactic haloes.

We compare the results in Fig.~\ref{Fig:PS_3cols} with the phase portraits of filaments and knots of \cite{Martizzi2019a}, whose results also derive from an analysis of the IllustrisTNG simulation (see their Fig. 4). 
In that paper, the phase portrait of filaments exhibits a large number of cells aligned in the elliptical pattern (from WHIM to Hot medium) specific to collapsed structures, which makes it slightly different from our findings here. Indeed, we will see in Sect.~\ref{Sect:ContributionGalaxies} that the elliptical pattern in the phase-space of filaments in \cite{Martizzi2019a} is due to 
gas associated with haloes in filaments.
Moreover, the phase portrait of knots in their paper shows a mixture of contributions from both the galactic haloes, and the gas around the CPmax point identified in this work.
These differences can be explained by the very different cosmic classification method of \cite{Martizzi2019a}, where the association of gas to the different structures of the cosmic web (knots, sheets, filaments and voids) is performed with an algorithm based on the Hessian of the density field. On the contrary, in this paper, we detect filaments using the DisPerSE algorithm and, since these are defined as the ridges of the Delaunay tessellation (see Sect.~\ref{SubSect:fil_cata}), we are able to  locate the position of the filament spines as well as those of the CPmax points. Therefore, our results are an interesting complement to those of \cite{Martizzi2019a}, as we have identified galactic halo signatures in the portraits of filaments and nodes, and we have shown that the gas content of different populations of filaments is not the same.


\section{\label{Sect:Results2}Thermodynamical properties}

In the previous section, we focused on the distribution of the different gas phases around the filaments of the cosmic web, and we analysed the gas content of these structures by performing phase portraits of the gas cells. We will now study a couple of thermodynamical properties of gas, namely the temperature and the pressure, as a function to the distance to the filament. How do these quantities vary as we get closer to the filament spine, starting from the outskirts? Do the different gas phases described in Sect.~\ref{SubSect:gas_phases_definition} vary in the same way? How does the picture change form one type of filaments to the other?

\subsection{\label{SubSect:method_temperature_profiles}Method}

We build radial profiles of gas temperature and pressure as functions of the distance to the spine of the filament.
These are computed as volume-weighted averages of gas cell quantities, as shown by Eq.~\ref{Eq:Tprofile} and ~\ref{Eq:Pprofile}, respectively for temperature and pressure:

\begin{equation}\label{Eq:Tprofile}
    T(r_k) = \frac{\displaystyle
    \sum_{s = 1}^{N_\mathrm{seg}}
   \left( \sum_{j = 1}^{N(k)} T_{\mathrm{cell}, j} \times V_{\mathrm{cell}, j} \right)_{s}
    }{
    \displaystyle \sum_{s = 1}^{N_\mathrm{seg}} \left( \sum_{j = 1}^{N(k)} V_{\mathrm{cell}, j } \right)_{s} 
    } \ \ ,
\end{equation}

\begin{equation}\label{Eq:Pprofile}
    P(r_k) = \frac{\displaystyle
    \sum_{s = 1}^{N_\mathrm{seg}}
   \left( \sum_{j = 1}^{N(k)} P_{\mathrm{cell}, j} \times V_{\mathrm{cell}, j} \right)_{s}
    }{
    \displaystyle \sum_{s = 1}^{N_\mathrm{seg}} \left( \sum_{j = 1}^{N(k)} V_{\mathrm{cell}, j } \right)_{s} 
    } \ \ .
\end{equation}
At the distance $r_k$ from the spine of the filament, we sum the products between temperature and volume ($T_\mathrm{cell} \times V_\mathrm{cell}$) of the $N(k)$ cells inside the $k$-th cylindrical shell around the segment $s$. We repeat this procedure for the same $k$-th cylindrical shell of all the segments and we sum all contributions. We divide this result by the sum of the volumes of all cells in the corresponding $k$-th shell of each segment. This method\footnote{We note that for a study of combined quantities (e.g. mean $T \times P$), a simple multiplication of individual profiles would be incorrect, since one would need to apply again Eq.~\ref{Eq:Tprofile} to the combination of variables at the cell level.} is also applied to the pressure profiles (Eq.~\ref{Eq:Pprofile}).
By computing volume-weighted profiles we retrieve results that are independent of the Arepo grid, i.e. independent of the irregular volumes of the simulation's Voronoi gas cells. More explicitly, the cell refinement criterion of the moving-mesh Arepo code is based on a fixed mass threshold \citep{Weinberger2020arepo}, leading to a broad distribution of cell volumes. Therefore, the more conventional mass-weighted average profiles would have been biased by the different volumes of the gas cells. Finally, we specify that the error bars in all the following plots are estimated using the bootstrap method applied on the distribution of segments, as described in Sect.~\ref{SubSect:gas_fraction_profiles}.

\subsection{\label{SubSect:T_profiles}Temperature profiles}

    \begin{figure}
   \centering
   \includegraphics[width=0.5\textwidth]{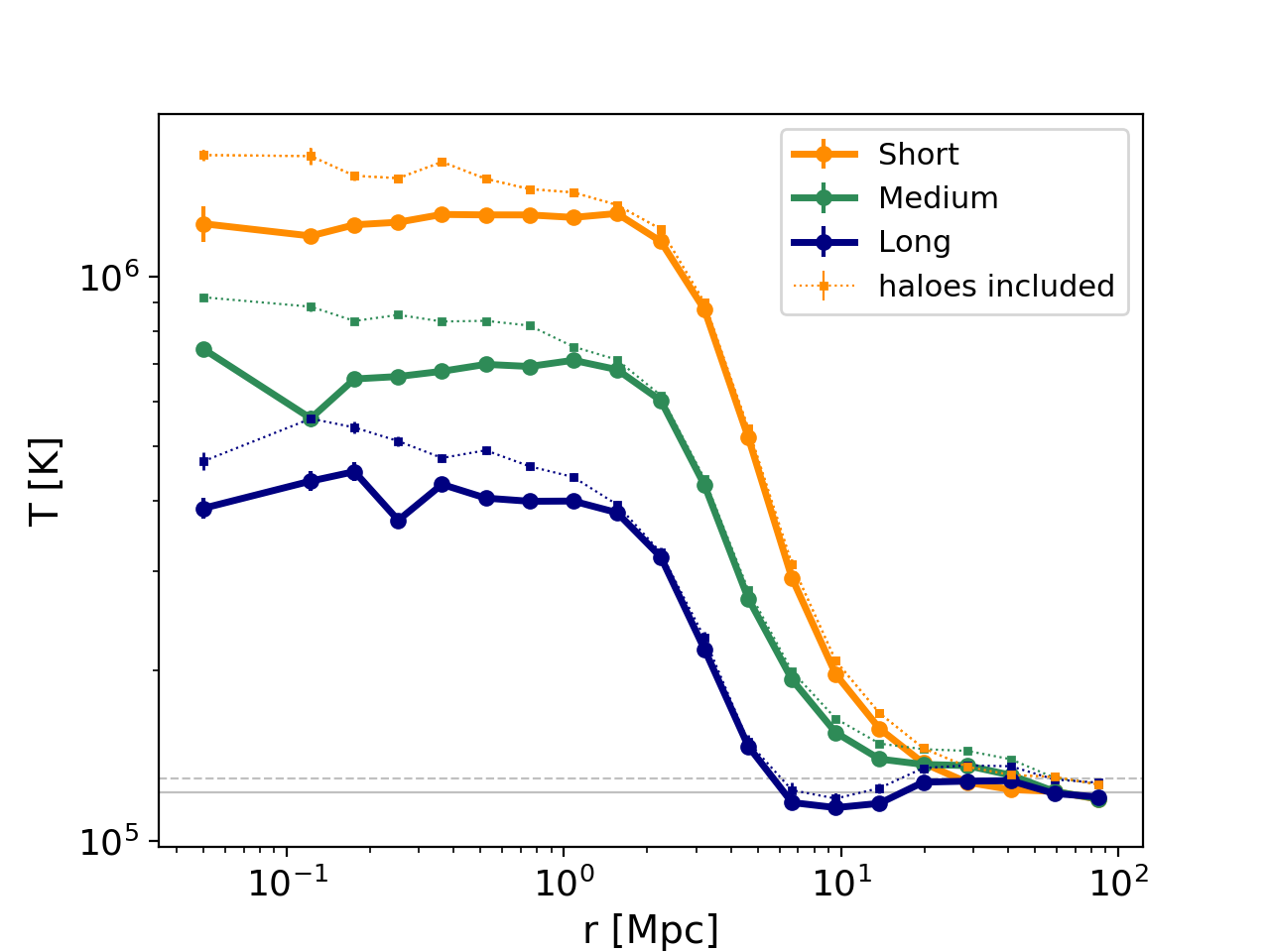}
   \caption{Temperature profiles of short ($L_f < 9$ Mpc), medium-length ($9 \leq L_f < 20$ Mpc), and long filaments ($L_f \geq 20$ Mpc) respectively in orange, green and blue colours. Thick solid lines with circles correspond to the temperature profiles presented in Sect.~\ref{SubSect:T_profiles}, while thin dotted lines with squares show the results where contributions of galactic haloes are included (see Sect.~\ref{Sect:ContributionGalaxies}). For each case, the horizontal gray lines represent the average temperatures of all the gas cells (CPmax points excluded) in the box. These profiles clearly show the isothermal cores of filaments.}
    \label{Fig:Temp_pro_NoPhases}
    \end{figure}

Let us first focus on the total temperature profiles of gas around filaments, regardless of the gas phase. Figure~\ref{Fig:Temp_pro_NoPhases} shows the temperature profiles for short ($L_f < 9$ Mpc), medium-length ($9 \leq L_f < 20$ Mpc), and long filaments ($L_f \geq 20$ Mpc) respectively in orange, green and blue.
We focus here on the thick solid lines with circles, given that the thin dotted lines with squares correspond to the results when contributions of galactic haloes are included, which will be discussed in the dedicated Sect.~\ref{Sect:ContributionGalaxies}.
The horizontal gray lines represent the mean temperatures of all the gas cells in the box (CPmax excluded), in each case.

These temperature profiles exhibit the main feature of an isothermal core up to $r_\mathrm{core} = 1.5$ Mpc from the axis of the filament. This trend is present in profiles of all three filament types (short, medium-length and long). It is worth noticing that the radial scale of $r_\mathrm{core} = 1.5$ Mpc is independent from the resolution of the simulation, as shown in Appendix~\ref{Appendix:Resolution}. Most likely, a more fundamental physical reason is behind this characteristic scale (for example, an equilibrium between the gas pressure, volume and density), but any conclusion is still premature at this stage and would require further investigations that go beyond the scope of the present paper. We note, however, that isothermal filament cores have  already been observed in the temperature profiles of single simulated filaments in e.g. \cite{KlarMucket2012} and \cite{GhellerVazza2019_surveyTandNTprops_fils}, and here we retrieve this property in a statistically significant way. We note that in the latter paper, the authors use an algorithm based on fixed density thresholds to detect filaments in the gas density field, which is a very different approach from the DisPerSE algorithm employed here.

Interestingly, we observe that the value of the plateau, $T_\mathrm{core}$, strongly depends on the length of the filament (e.g. orange vs blue curves). For quantitative purposes, Table~\ref{Table:TPtable} reports the different values of $T_\mathrm{core}$, computed directly from the profiles as the average temperature of the points at $r < 1.5$ Mpc.
In the left side of this table, we clearly see that cores of short filaments are more than three times hotter than those of long filaments. Indeed, the gravitational heating efficiency is likely stronger in the short population, due to the deeper potential wells of these denser regions. This result is independent of the presence or the absence of galactic haloes, which will be discussed in Sect.~\ref{Sect:ContributionGalaxies}. We note that the temperature ranges of cores of filaments found in our analysis are in agreement with the recent work of \cite{Tuominen2020}, that studied filaments detected with the Bisous algorithm \citep{Tempel2016bisous} in the EAGLE simulation \citep{Schaye2015_EAGLEsimu}.

At distances larger than $r = 1.5$ Mpc from the filament spine, the gas average temperature drops sharply. This drop is monotonic in short filaments, where the minimum of $T \sim 1.1 \times 10^5$ K is reached in the most distant regions from the filament ($>50$ Mpc). We emphasise that this value represents the average temperature in regions that are very far away from the spine, so gas from other structures (namely other, distant filaments) is included in this average. We checked that, by masking the gas cells from other filaments (at $r \leq 2$ Mpc from the spine), we remove their contribution to the temperature and we find a lower background average, that is $T \sim 8.5 \times 10^4$ K.
Unlike for short filaments, the temperature profiles of the long filaments do not follow a monotonic trend. They exhibit a global minimum at $r \sim 10$ Mpc, followed by a slight increase of the temperature, to finally reach the same background level as short (and medium-length) filaments. 
These dissimilarities between populations of short and long filaments at distances larger than $r = 1.5$ Mpc might come from their different environments in the cosmic web. Indeed, since short filaments statistically reside in denser regions \citep{GalarragaEspinosa2020}, they are more likely to be closely surrounded by other dense structures (e.g. other filaments, clusters or haloes) that contribute to the relatively high average temperature on the $r > 1.5$ Mpc outskirts. On the contrary, long filaments, tracers of less dense environments, extend into under-dense regions (voids) where the gas is cooler, which explains the dip at $r \sim 10$ Mpc, before the temperature reaches the average value of the entire box. Note that these differences in temperature between short and long filaments reflect well with the phase portraits of Fig.~\ref{Fig:PS_3cols}.\\

\begin{figure*}
   \centering
   \includegraphics[width=0.35\textwidth]{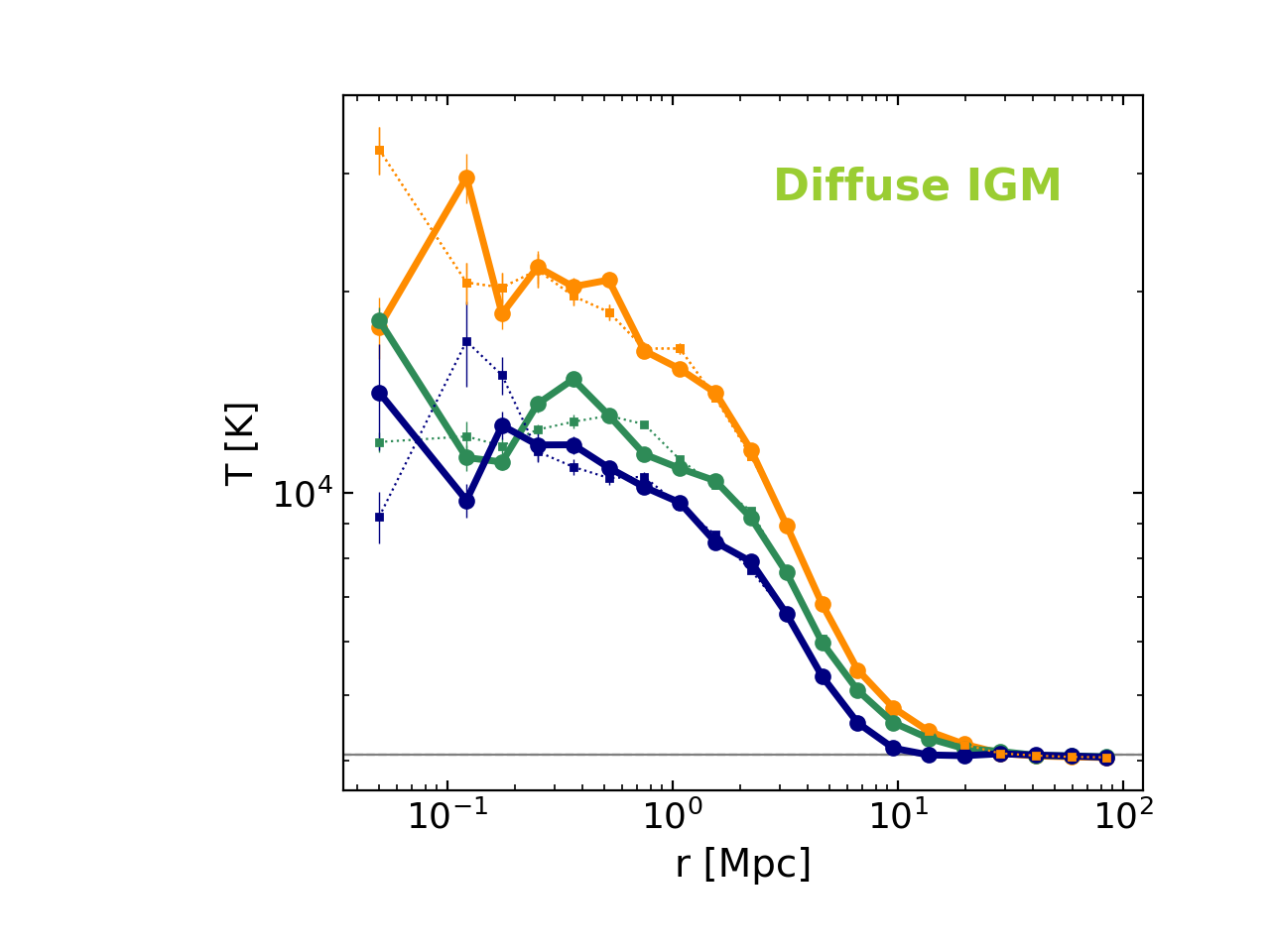}\includegraphics[width=0.35\textwidth]{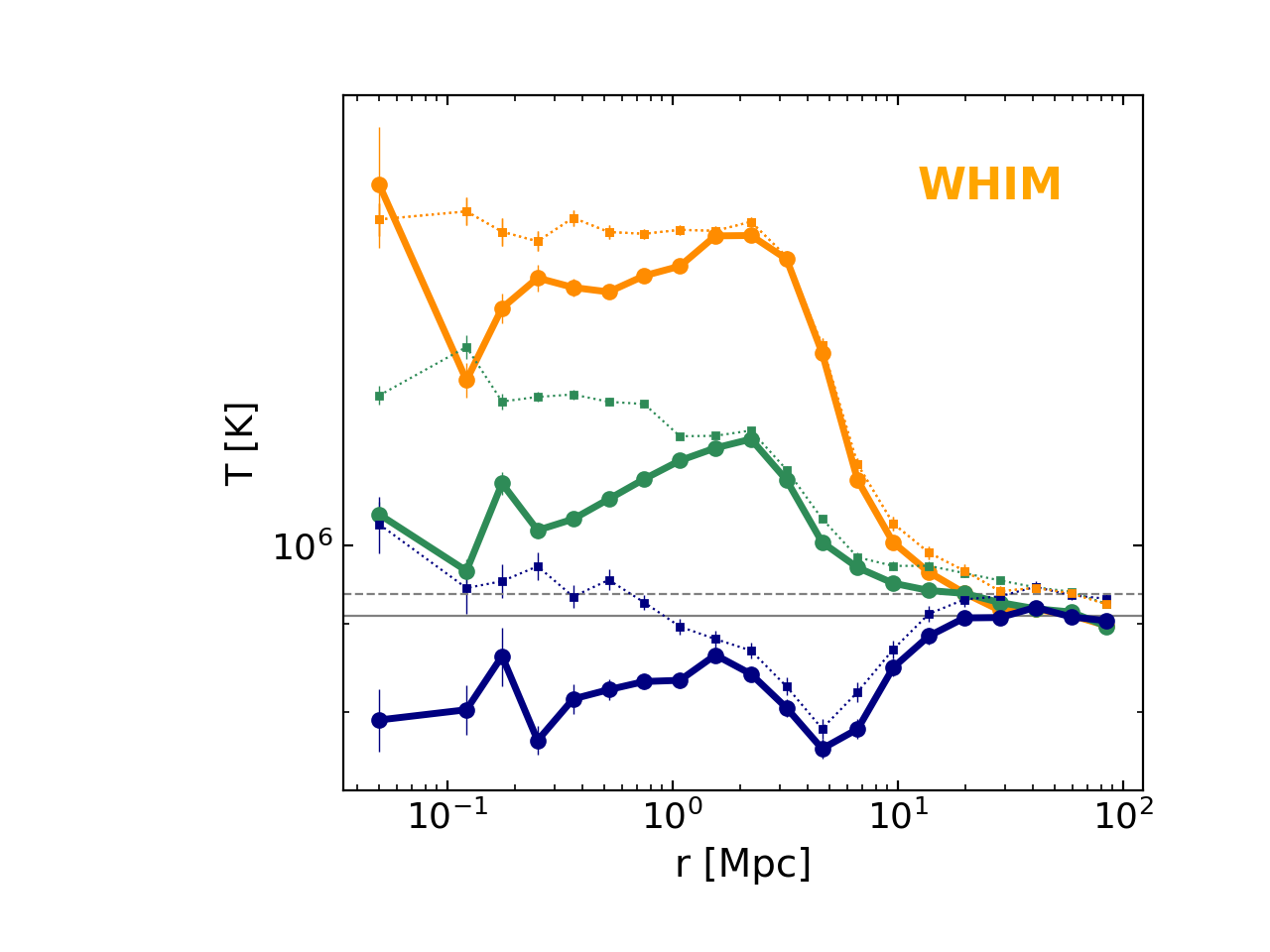}\includegraphics[width=0.35\textwidth]{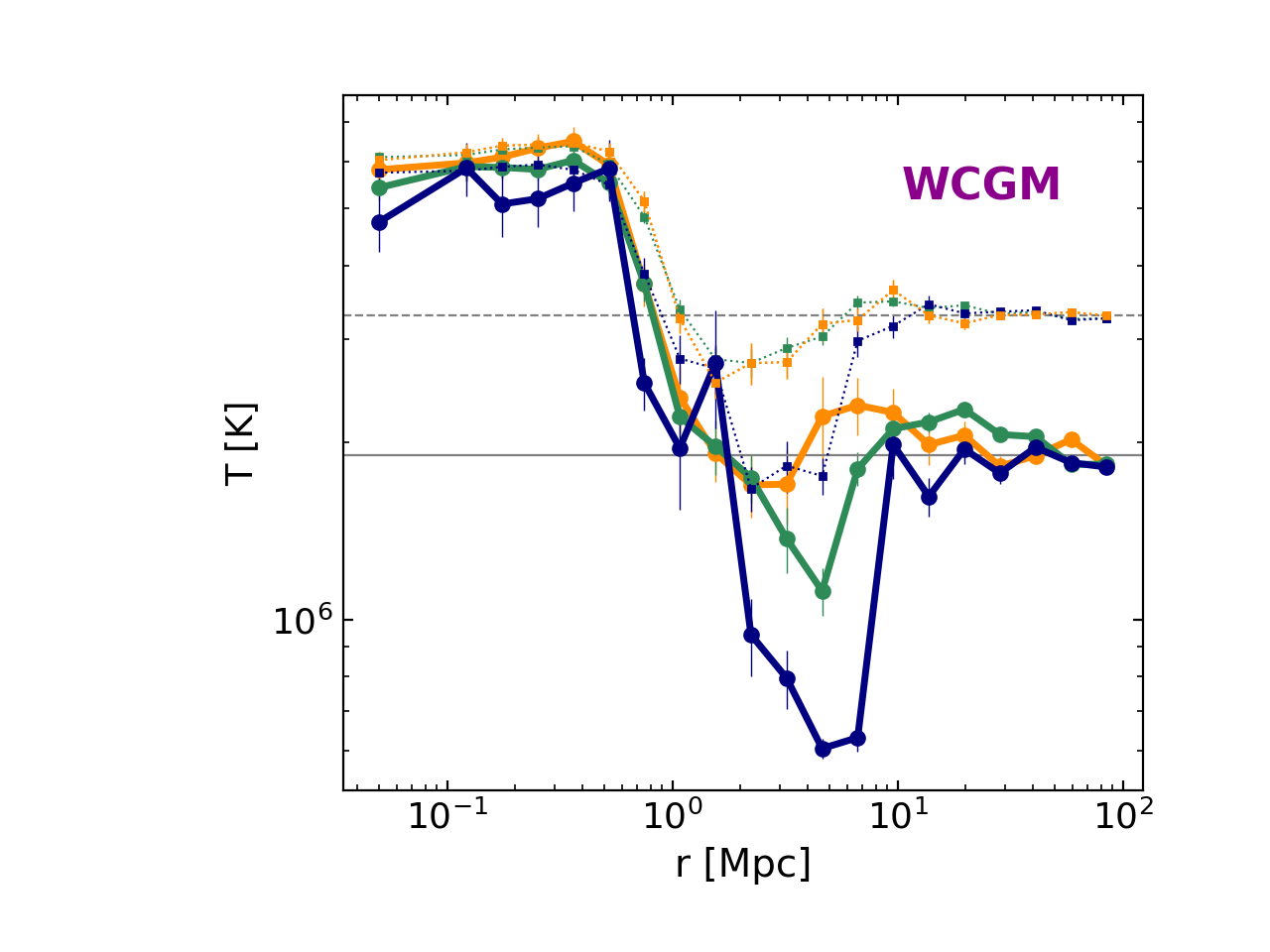}
   \includegraphics[width=0.35\textwidth]{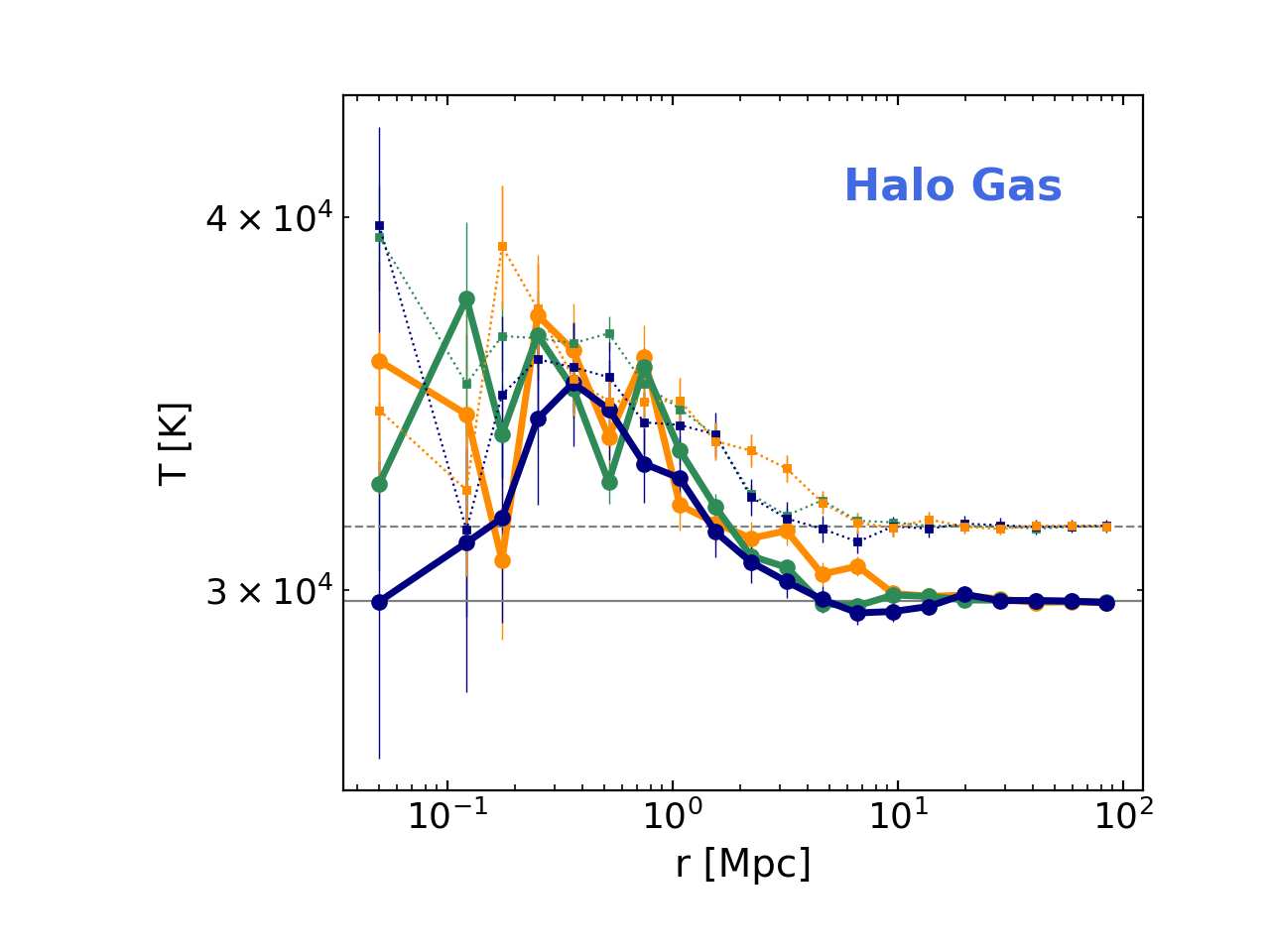}\includegraphics[width=0.35\textwidth]{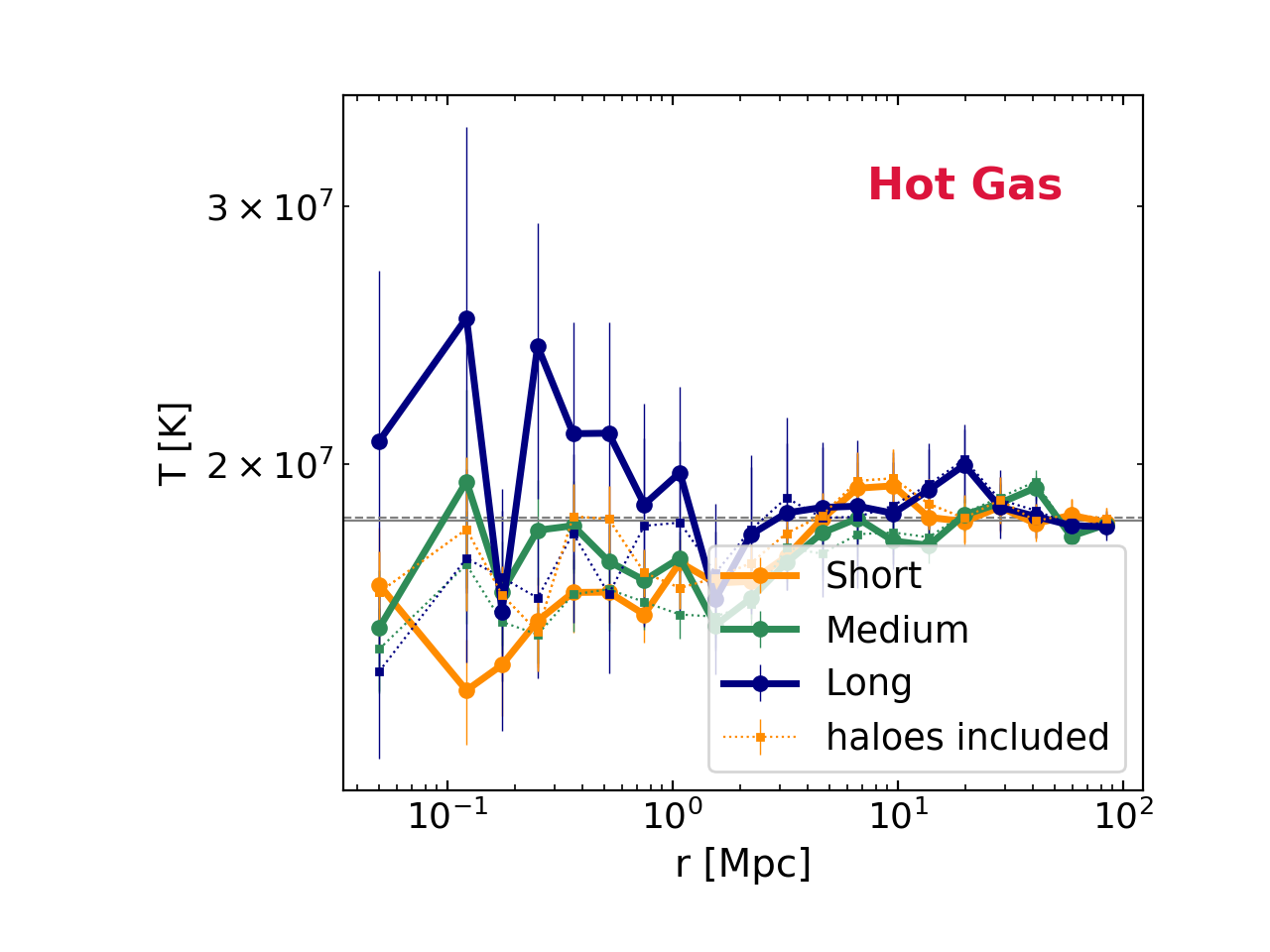}
   \caption{Temperature profiles of the different gas phases (see Table~\ref{Table:gas_phases}), for short ($L_f < 9$ Mpc, orange curves), medium-length ($9 \leq L_f < 20$ Mpc, green curves), and long filaments ($L_f \geq 20$ Mpc, blue curves). Thick solid lines with circles correspond to the temperature profiles presented in Sect.~\ref{SubSect:T_profiles}, while thin dotted lines with squares show the results where contributions of galactic haloes are included (see Sect.~\ref{Sect:ContributionGalaxies}). For each case, the horizontal gray lines represent the average temperatures of all the gas cells (CPmax points excluded) in the box.}
    \label{Fig:Temp_pro}
    \end{figure*}

Let us now split the different gas phases to see how their temperatures behave around filaments. We compute the temperature profiles of the five different phases, and we present them in the panels of Fig.~\ref{Fig:Temp_pro}, where the same colour code as in Fig.~\ref{Fig:Temp_pro_NoPhases} is adopted. In this section, we focus on the thick solid lines with circles, and the thin dotted lines with squares will be discussed in Sect.~\ref{Sect:ContributionGalaxies}.
We note that these profiles span a very broad total temperature range, from cold temperatures of $T \sim 10^{4}$ K exhibited by the Diffuse IGM (upper right panel), to the very hot values, of the order of $T \sim 2 \times 10^{7}$ K, reached by the Hot gas (lower right panel). This broad range simply reflects the definition of the gas phases.

The temperature profiles of the five gas phases in Fig.~\ref{Fig:Temp_pro} exhibit very different shapes and behaviours depending on the type of filament. All the gas phases around the short population show a rise in temperature from the outskirts to the core, except the Hot Gas that remains constant. The same trend is followed by gas around medium-length filaments, but in this case the temperature rise of the profile of WHIM is only very mild. Finally, the temperature profiles of the phases around long filaments exhibit a broader diversity.

The fact that the temperature profile of the Hot gas phase remains essentially flat is easily understood by looking at Fig.~\ref{Fig:PS_3cols}, where the vast majority of the gas cells in filaments occupy only a limited and rather flat region that extends into the Hot gas phase domain. Moreover, our results are in agreement with recent findings from X-ray ROSAT observations. Indeed, \cite{Tanimura2020_Rosat} detected a significant X-ray emission from hot gas at temperatures of $1.0^{+1.1}_{-0.7} \times 10^7$ K ($= 0.9^{+1.0}_{-0.6}$ keV) in regions of cores of filaments, and this range is compatible with the mean temperature values of hot gas presented in this work.

To the contrary of all phases, Diffuse IGM gas (upper right panel) exhibits the most significant temperature increase, from the background to the core of filaments. For example, cores of short filaments are up to seven times hotter than their background temperature. The increase of temperature from background to core is smaller in long filaments (for which the temperature is only multiplied by three), and this might be related to their location in less dense environments of the cosmic web \citep[see][, and references therein]{GalarragaEspinosa2020}. For all filaments, this significant temperature increase in the Diffuse IGM profile may be due to the gravitational heating resulting from the accretion of this diffuse gas into the filament (as discussed in Sect.~\ref{SubSect:gas_fraction_profiles}). We note that the maximum temperature is not higher than $10^5$ K, given that beyond this threshold gas is counted as part of the WHIM phase.

Concerning the WHIM (middle left panel), we observe notably different temperature profiles for short, medium-length and long filaments. The temperature of WHIM gas around short filaments increases with decreasing distance to the core, while in long filaments, the WHIM temperature shows a decrease on the outskirts ($r \sim 6-10$ Mpc). Once again, this can be seen as a consequence of the different cosmic web environments of the short and long filaments. Gas accreted towards short filaments is likely rapidly heated by shocks, mergers and feedback effects taking place in the denser environments that characterise these structures. On the contrary, gas falling into long filaments is probably not subject to the same dynamics (or perhaps simply with lesser efficiency) as gas falling into the short population.

Interestingly enough, the temperature profiles of the two phases corresponding to gas in and around haloes, respectively Halo gas and WCGM, do not appear to be sensitive to the different types of filaments (and therefore to the large scale environment). As we move towards the spine, circumgalactic gas gas gets heated by shocks and feedback processes near galaxies in fairly the same way for short and long filaments, as shown by their very similar profiles (see coloured curves in the WCGM and Halo gas panels of Fig.~\ref{Fig:Temp_pro}). 
The profiles of these phases rather present a notable difference in the mean temperature levels (gray horizontal lines) weather galactic haloes are included (dotted lines) into the temperature estimation or not (solid lines). This feature is expected, given that Halo gas and WCGM are tracers of haloes, and that these collapsed structures can be located in background and foreground filaments, thus contributing to the rise of the mean temperature. Further details on the part that galactic haloes contribute to the temperature of filaments will be discussed in Sect.~\ref{Sect:ContributionGalaxies}.

\subsection{\label{SubSect:P_profiles}Pressure profiles}

    \begin{figure}
   \centering
   \includegraphics[width=0.5\textwidth]{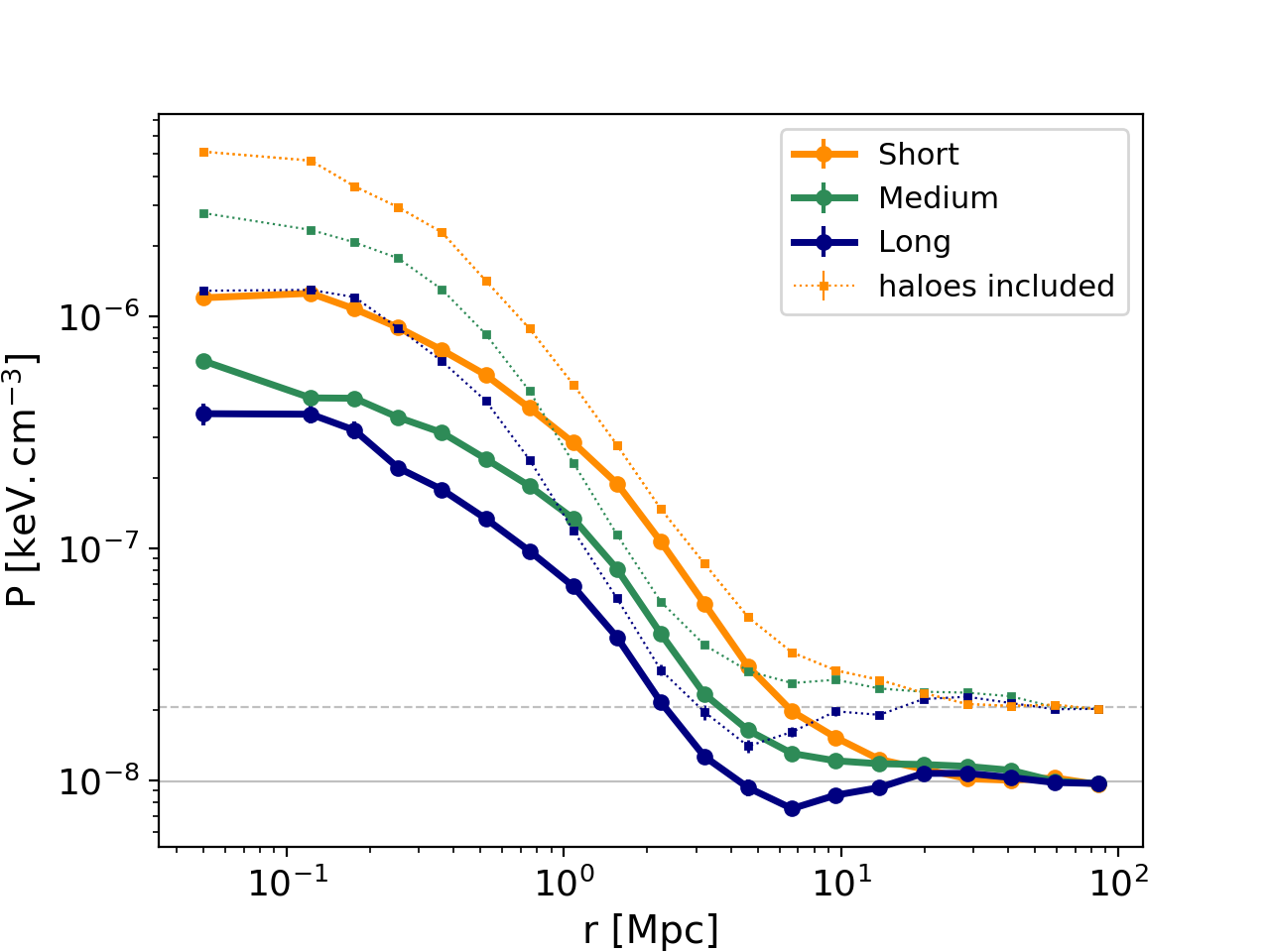}
   \caption{Pressure profiles of short ($L_f < 9$ Mpc, orange curves), medium-length ($9 \leq L_f < 20$ Mpc, green curves), and long filaments ($L_f \geq 20$ Mpc, blue curves). Thick solid lines with circles correspond to the pressure profiles presented in Sect.~\ref{SubSect:T_profiles}, while thin dotted lines with squares show the results where contributions of galactic haloes are included (see Sect.~\ref{Sect:ContributionGalaxies}). For each case, the horizontal gray lines represent the average pressures of all the gas cells (CPmax points excluded) in the box.}
    \label{Fig:Pres_pro_NoPhases}
    \end{figure}

   \begin{table*}
    \caption{Temperature $T_\mathrm{core}$ and pressure $P_\mathrm{core}$ at the cores of short ($L_f < 9$ Mpc), medium-length ($9 \leq L_f < 20$ Mpc), and long filaments ($L_f \geq 20$ Mpc). The temperature values are computed as the average temperature of the points at $r < 1.5$ Mpc, while pressure corresponds to the value at the first radial bin, i.e. $r=0.05$ Mpc.}
    \label{Table:TPtable}     
    \centering  
    \begin{tabular}{ l | c  c | c  c }
    \hline\hline 
    & \multicolumn{2}{c|}{Inter-filament gas} & \multicolumn{2}{c}{Haloes included} \\ 
    & $T_\mathrm{core}$ [K] & $P_\mathrm{core}$ [$\mathrm{keV.cm}^{-3}$] & $T_\mathrm{core}$ [K] & $P_\mathrm{core}$ [$\mathrm{keV.cm}^{-3}$] \\ \hline
     Short & $1.3 \times 10^6$ & $1.2 \times 10^{-6}$ & $1.5 \times 10^6$ & $5.1 \times 10^{-6}$ \\
     Medium & $6.8 \times 10^5$ & $6.4 \times 10^{-7}$ & $8.4 \times 10^5$ & $2.8 \times 10^{-6}$ \\
     Long &  $4.1 \times 10^5$ &  $3.8 \times 10^{-7}$ &  $4.9 \times 10^5$ &  $1.3 \times 10^{-6}$ \\
     \hline
    \end{tabular}
    \end{table*}

We compute the profiles of gas pressure (see Sect.~\ref{SubSect:cell_TandP_method}) around filaments using the same method as for temperature, described in Sect.~\ref{SubSect:method_temperature_profiles}. Before showing the results for each gas phase, let us first focus on the general pressure profiles computed considering all the gas cells.

The latter are shown in Fig.~\ref{Fig:Pres_pro_NoPhases} for short, medium-length and long filaments.
As expected from our previous results, filaments of different lengths have significantly different pressure profiles. Indeed, at the core ($r \leq 1$ Mpc) and on the outskirts ($1 < r \leq 10$ Mpc), short filaments show values that are more than three times those of long filaments, as we reported in the left side of Table~\ref{Table:TPtable} (where the values correspond to the pressure read from the first radial bin). The same trends as for the temperature profiles of Fig.~\ref{Fig:Temp_pro_NoPhases} are observed in Fig.~\ref{Fig:Pres_pro_NoPhases}, that are the monotonic decrease of short filament profiles, and the presence of a global minimum in the pressure on the outskirts of long filaments. Again, these differences might be related to the different environments traced by the two populations (see text in Sect.~\ref{SubSect:T_profiles}).
In comparison with the findings of \citet{Arnaud2010} who showed that pressure in the cores of galaxy clusters lies in the range of $P = 1-300 \times 10^{-3}$ $\mathrm{keV.cm}^{-3}$, we find that pressure in filaments, $P = 4-12 \times 10^{-7}$ $\mathrm{keV.cm}^{-3}$, is more than $1000$ times smaller. These three orders of magnitude difference is not surprising, given the fact that the galaxy clusters are the densest and hottest structures of the cosmic web.\\

\begin{figure*}
   \centering
  \includegraphics[width=0.35\textwidth]{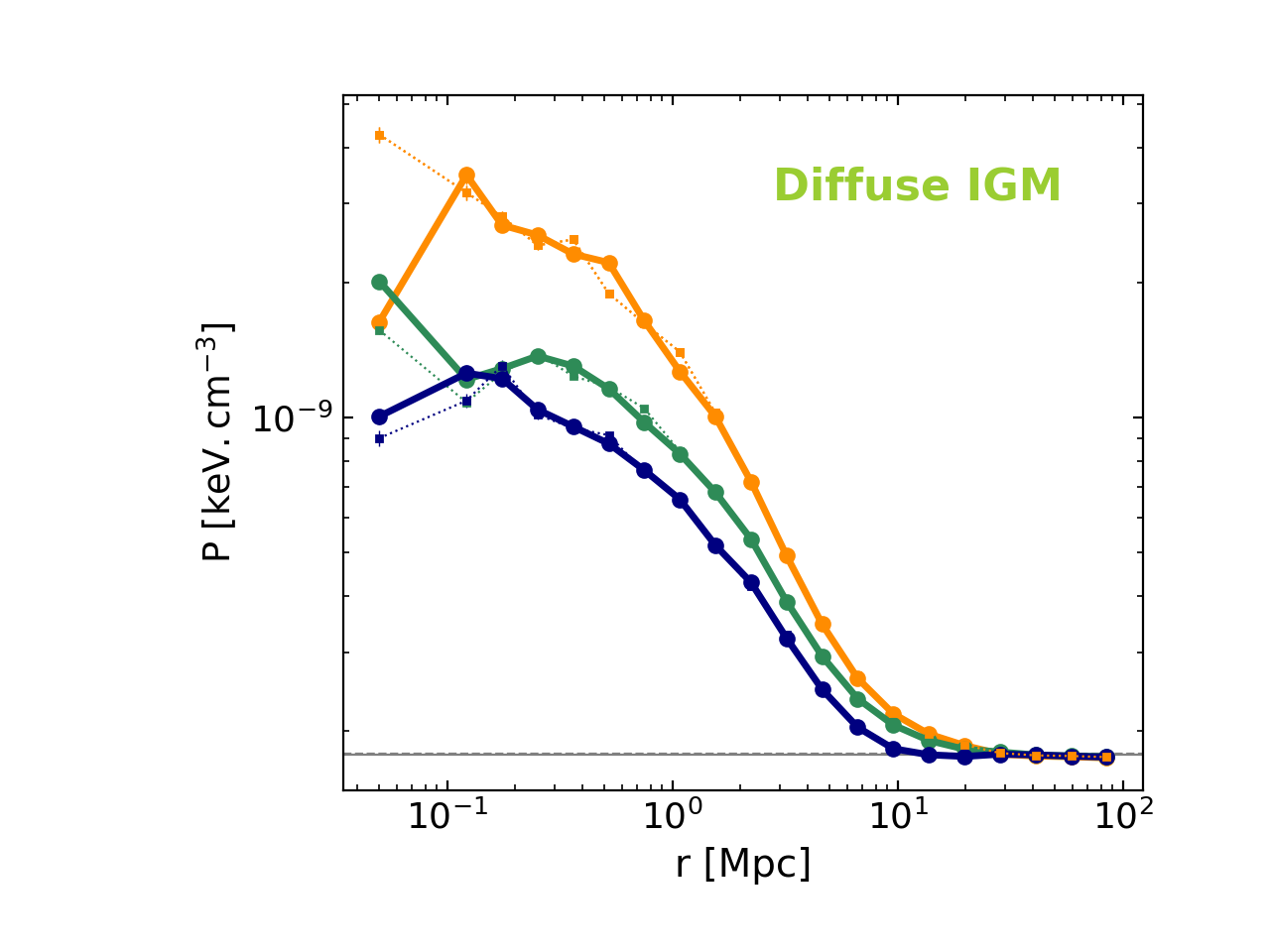}\includegraphics[width=0.35\textwidth]{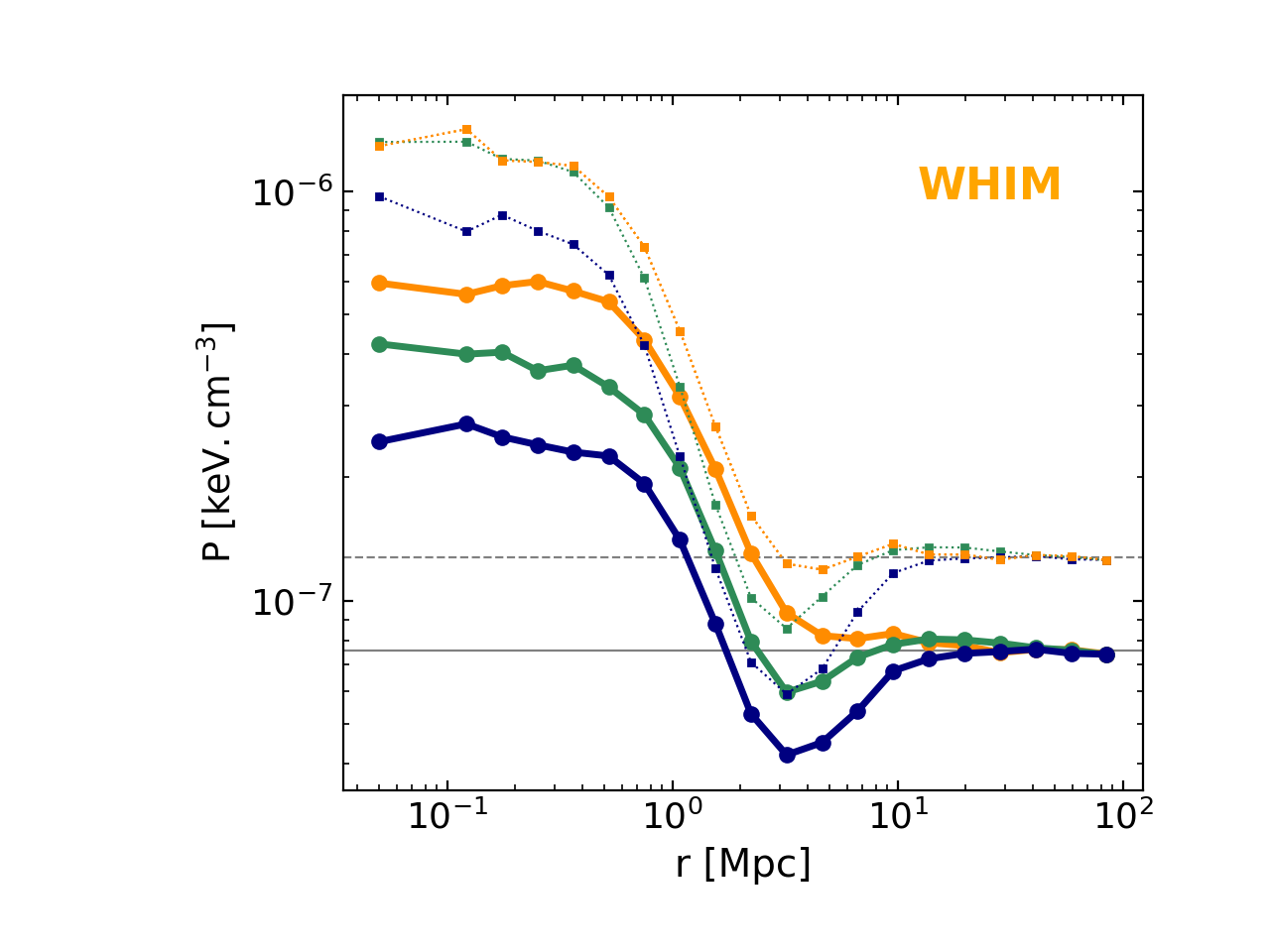}\includegraphics[width=0.35\textwidth]{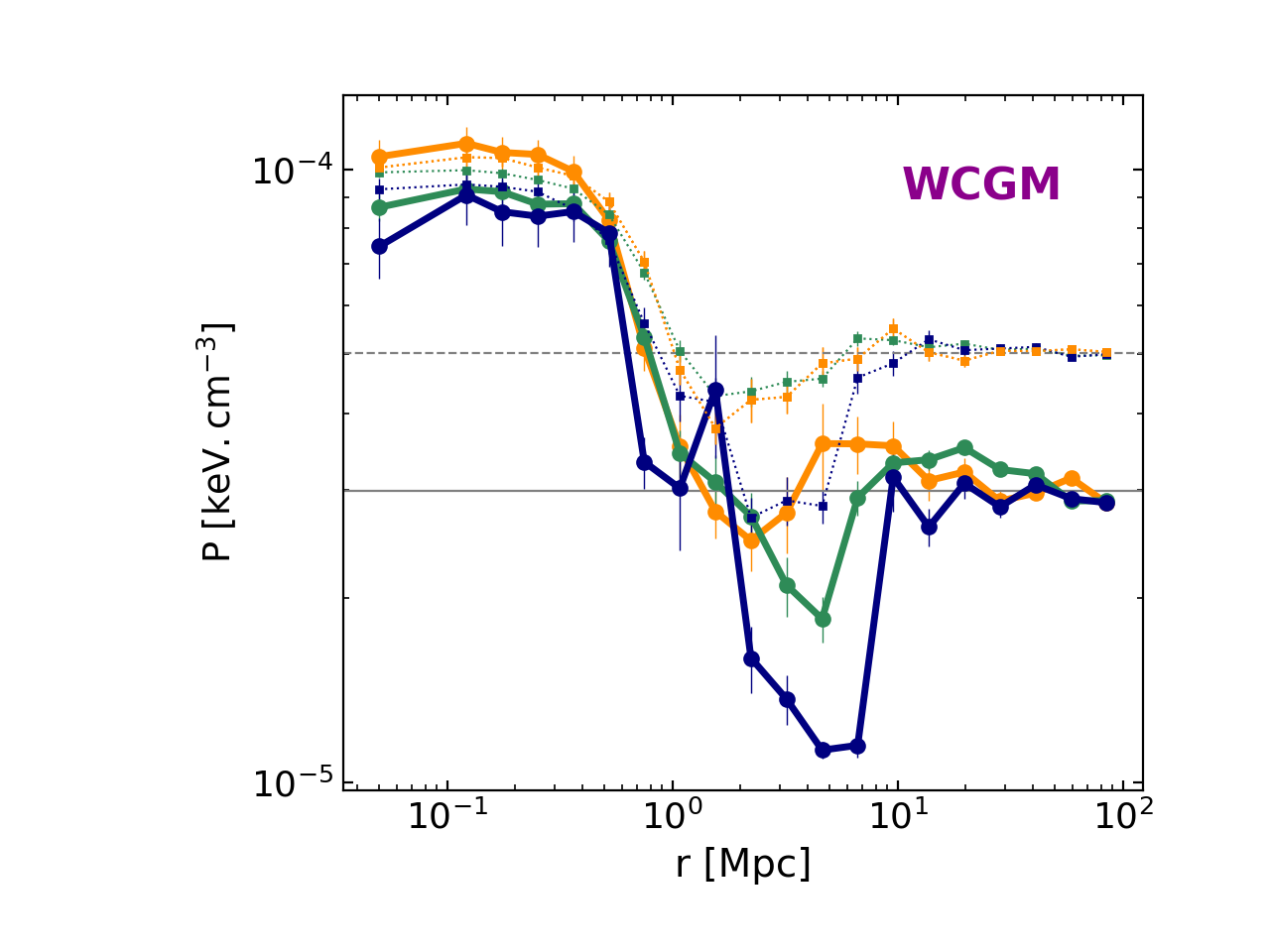}
  \includegraphics[width=0.35\textwidth]{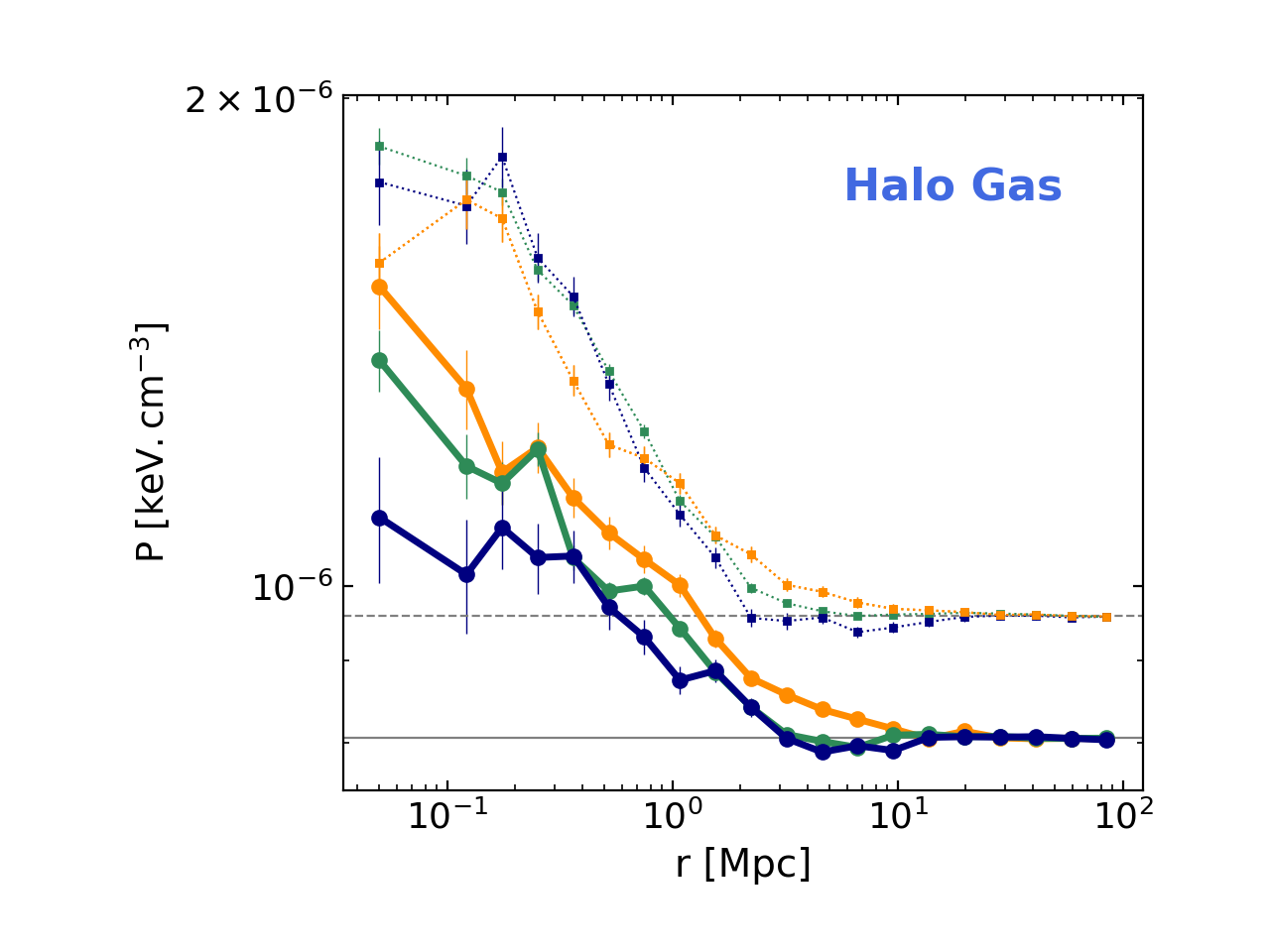}\includegraphics[width=0.35\textwidth]{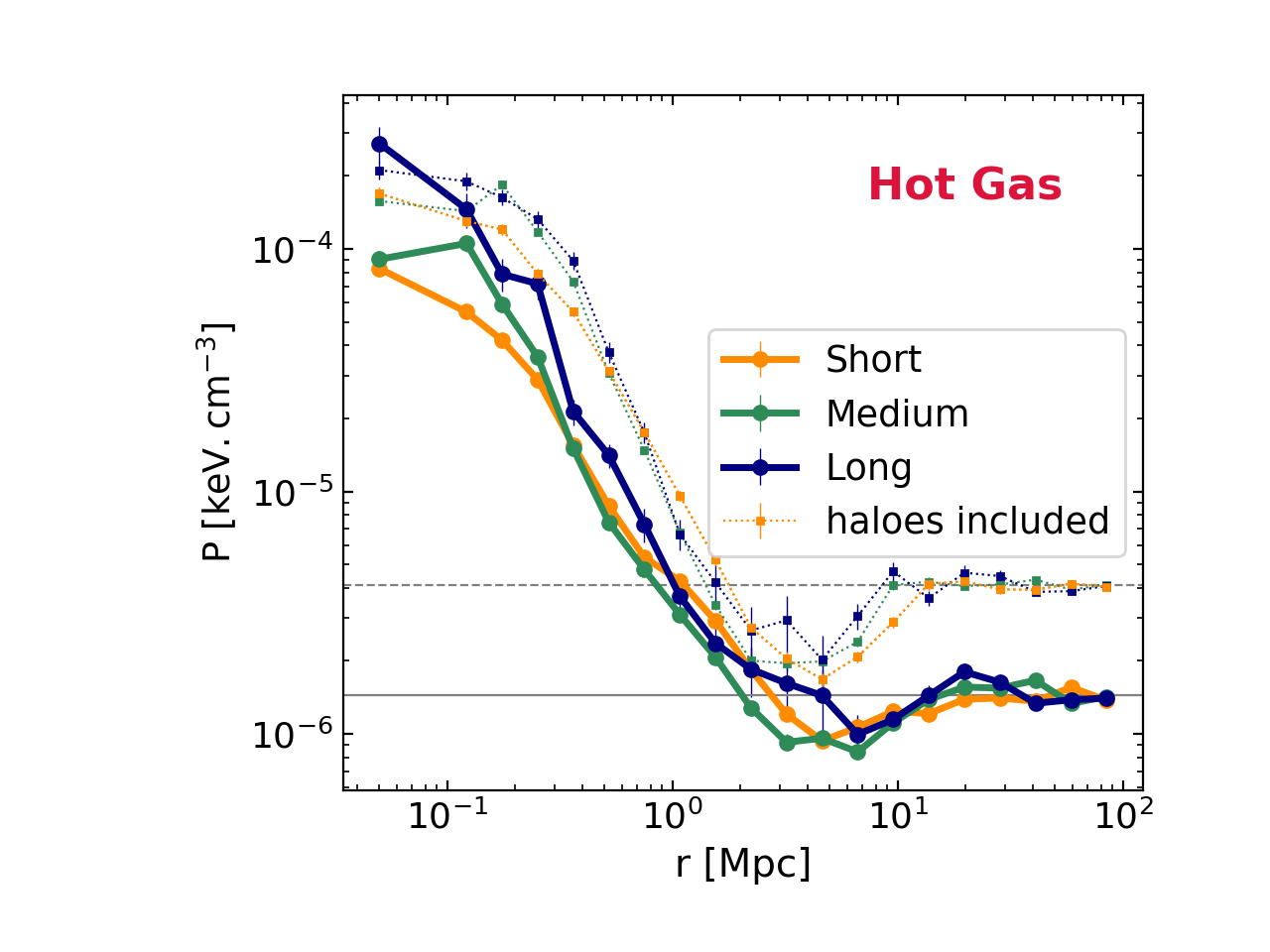}
   \caption{Pressure profiles of the different gas phases (introduced in Sect.~\ref{SubSect:gas_phases_definition}), for short ($L_f < 9$ Mpc, orange curves), medium-length ($9 \leq L_f < 20$ Mpc, green curves), and long filaments ($L_f \geq 20$ Mpc, blue curves). The pressure profiles in thick solid lines with circles are presented in Sect.~\ref{SubSect:P_profiles}, while the dotted lines with squares correspond to the results including galactic haloes (discussed in Sect.~\ref{Sect:ContributionGalaxies}). The gray horizontal lines represent the average pressure of gas in each case.}
    \label{Fig:Pres_pro}
    \end{figure*}

Finally in Fig.~\ref{Fig:Pres_pro}, we show the pressure profiles of each gas phase. We note, as for the temperature, the very broad and particular pressure ranges of the different phases. Notably, the WCGM and Hot phases reach maximum values of $P \sim 10^{-4}$ $\mathrm{keV.cm}^{-3}$, that are comparable to pressures on the outskirts of clusters of galaxies \citep{Arnaud2010}.
We note that the pressure profiles of these phases, along with those of Halo gas, do not show a strong dependency on the type of filament (i.e. the coloured curves are essentially the same), showing that the pressures of WCGM, Halo and Hot gas are quite insensitive to the large scale environment in the cosmic web (which is expected for phases that are associated with collapsed structures, like haloes). This will be further discussed in Sect.~\ref{Sect:ContributionGalaxies}.
The opposite trend is observed in the pressure profiles of Diffuse IGM and WHIM phases (first and second panels of Fig.~\ref{Fig:Pres_pro}). Indeed, they present significant differences for short, medium-length and long filaments, showing that the properties of these phases are mainly ruled by the different environments in the cosmic web (traced by short and long filaments). This dependencies were already pointed out by the Diffuse IGM and WHIM temperature profiles of Fig.~\ref{Fig:Temp_pro}.

Let us now specifically focus on the WHIM phase (middle upper panel of Fig.~\ref{Fig:Pres_pro}), as some interesting features are exhibited by its pressure profiles. First, we notice that WHIM pressures increase by almost two orders of magnitude, from the outskirts to the cores of filaments. 
Almost all the profiles show a low-pressure zone at $r \sim 3$ Mpc (with respect to the mean value), and this is most marked in the results of long filaments (which are tracers of cosmic regions with low and moderate densities). 
This dip was already visible in Figs.~\ref{Fig:Temp_pro_NoPhases} and \ref{Fig:Pres_pro_NoPhases}, and is probably due to the under-dense environment around the longer filaments, where the gas is cooler and less dense.
This decrease of pressure happens at the same distance ($r \sim 3$ Mpc) from the cores of all the filaments regardless of their length, exhibiting a characteristic radius of WHIM
pressure around filaments. This common feature shows that WHIM gas might be subjected to the same physical processes (gravity and baryonic effects) in all types of filaments (though not with the same efficiencies, since the pressure values are different). Moreover, WHIM gas pressure does not follow the distribution of galaxies: in the same simulations, we have shown that galaxies are statistically closer to cores of long filaments than to short, showing a characteristic radius of $\sim 3$ and $\sim 5$ Mpc, respectively \citep{GalarragaEspinosa2020}.

\section{\label{Sect:ContributionGalaxies} Contribution of galactic haloes in filaments}

All the above results present the properties of gas in the inter-filament medium, i.e. excluding the contribution of collapsed structures. In this section, we compare our above findings to those obtained when the galactic haloes (i.e. massive $M_\mathrm{tot} > 10^{12}$ $\mathrm{M}_\odot$ haloes containing at least one galaxy, see Sect.~\ref{SubSect:gas_catalogue_building}) are retained.

\begin{figure}
   \centering
   \includegraphics[width=0.45\textwidth]{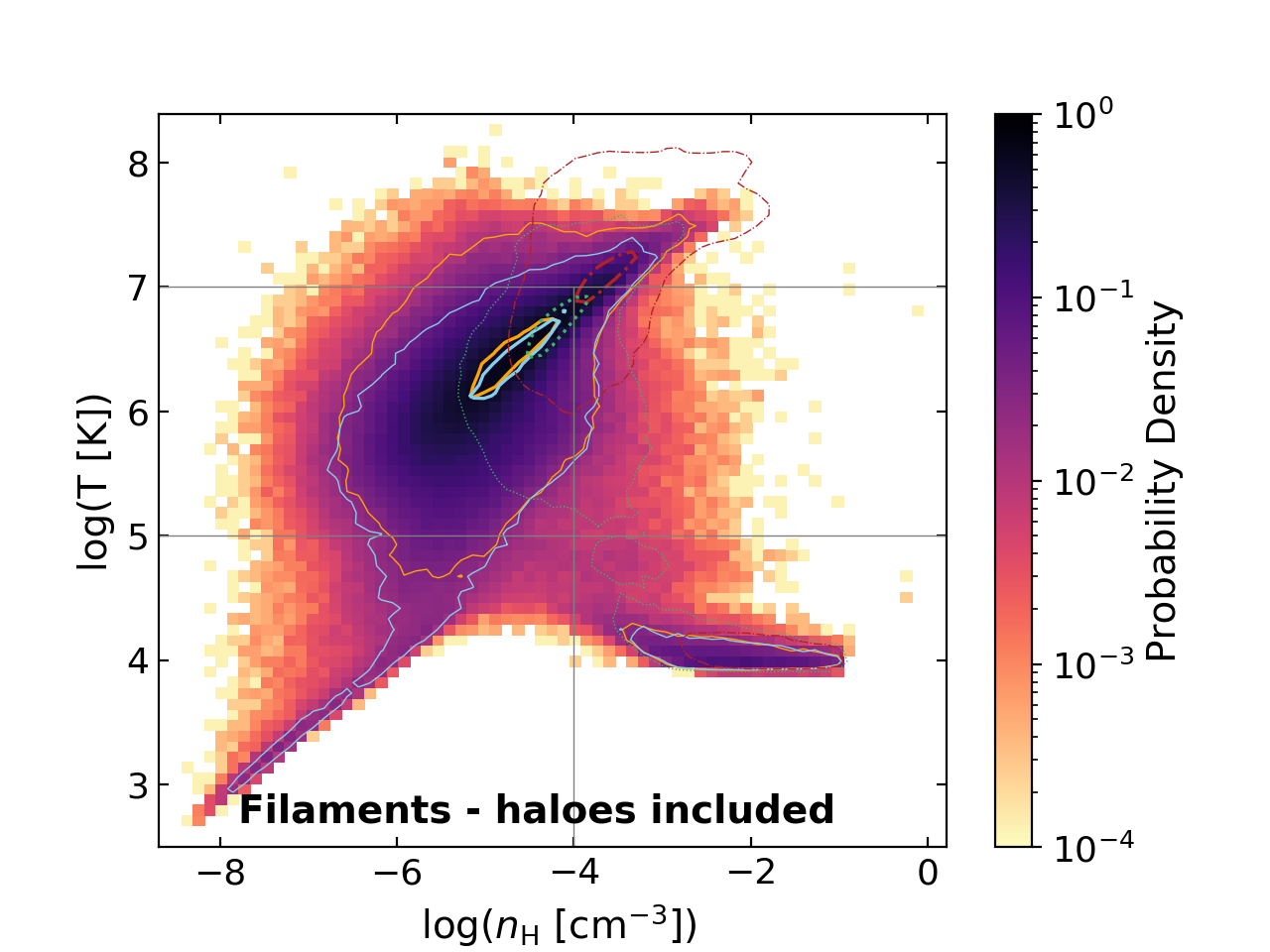}
   \caption{phase-space of gas in filaments, including the contribution of galactic haloes (i.e. haloes of mass $M_\mathrm{tot} > 10^{12} \, \mathrm{M}_\odot$ hosting at least one galaxy of mass $10^{9} \le \mathrm{M}_{*} [\mathrm{M_{\odot}}] \le 10^{12}$). The $68\%$ and $99\%$ contours correspond to results of short (orange) and long filaments (blue), to galactic haloes (dotted green lines), and gas around the maximum density critical points of the field (CPmax, in dashed-dotted red contours).}
    \label{Fig:PS_fils_haloesincluded}
    \end{figure}

Figure \ref{Fig:PS_fils_haloesincluded} presents the phase-space of gas within 1 Mpc from the spine of filaments, including the contribution from gas in galactic haloes residing in these core regions. We note substantial differences in the filament contours with respect to these of Fig.~\ref{Fig:PS_3cols}. Indeed, the contours are now in elliptical shapes that follow exactly those of galactic haloes (dotted green lines) towards the hottest and densest values at the boundaries between the WHIM, WCGM and Hot phases. The comparison of this phase-space to the first panel of Fig.~\ref{Fig:PS_3cols} shows explicitly the contribution of gas associated with galactic haloes residing in filaments. 
We note that Halo gas is also present in Fig.~\ref{Fig:PS_fils_haloesincluded}, with essentially the same significance as in Fig.~\ref{Fig:PS_3cols}, meaning that this phase cannot be interpreted as a contribution from massive galactic haloes (see Sect.~\ref{SubSect:Phase_Spaces}).

The contribution of galactic haloes on the temperature profiles of filaments (thin dotted lines with squares in Fig.~\ref{Fig:Temp_pro}) is a general rise of the temperature at the core. We report these core temperature values in the right side of Table~\ref{Table:TPtable}, which shows that cores of filaments appear on average $\sim 1.2$ times hotter when gas in haloes is included in the computation. This rise is expected as we have seen, in the phase-spaces of Fig.~\ref{Fig:PS_3cols} and \ref{Fig:PS_fils_haloesincluded}, that galactic haloes (statistically located in cores of filaments) contribute with hotter and denser gas.

Regarding pressure, the shapes and amplitudes of the total profiles of Fig.~\ref{Fig:Pres_pro_NoPhases} seem to be extremely sensitive to the presence of galactic haloes in filaments. Indeed, the values at the filament cores including galactic haloes (thin dotted lines with squares) are also reported in the right side of Table~\ref{Table:TPtable}, and happen to be more than three times larger than core pressures when haloes are removed. These differences are not surprising, since galactic haloes and clusters are known to reach pressures that are several orders of magnitude higher \citep[$\sim 1000$, see e.g. ][]{Arnaud2010} than our findings in the intra-filament medium, thus raising the average values.\\

We also disentangle the contributions of galactic haloes in the pressure and temperature profiles of each of the five different gas phases (see thin dotted lines with squares in Fig.~\ref{Fig:Temp_pro} and \ref{Fig:Pres_pro}). 
For the WHIM, WCGM, Halo, and Hot gas, we see that their values of temperature and pressure are clearly sensitive to the presence of galactic haloes, as solid and dotted profiles are significantly different both in temperature and pressure (see Fig.~\ref{Fig:Temp_pro} and \ref{Fig:Pres_pro}).
The inverse trends are shown by the Diffuse IGM phase.
Its properties are dominated by the large scale environment (of short and long filaments), but its temperature and pressure profiles are essentially the same whether galactic haloes are included or not, which simply reflects that this phase is absent in galactic haloes.

\section{\label{Sect:SZ}Sunyaev-Zel'dovich signal of filaments}

    \begin{figure*}
   \centering
   \includegraphics[width=1\textwidth]{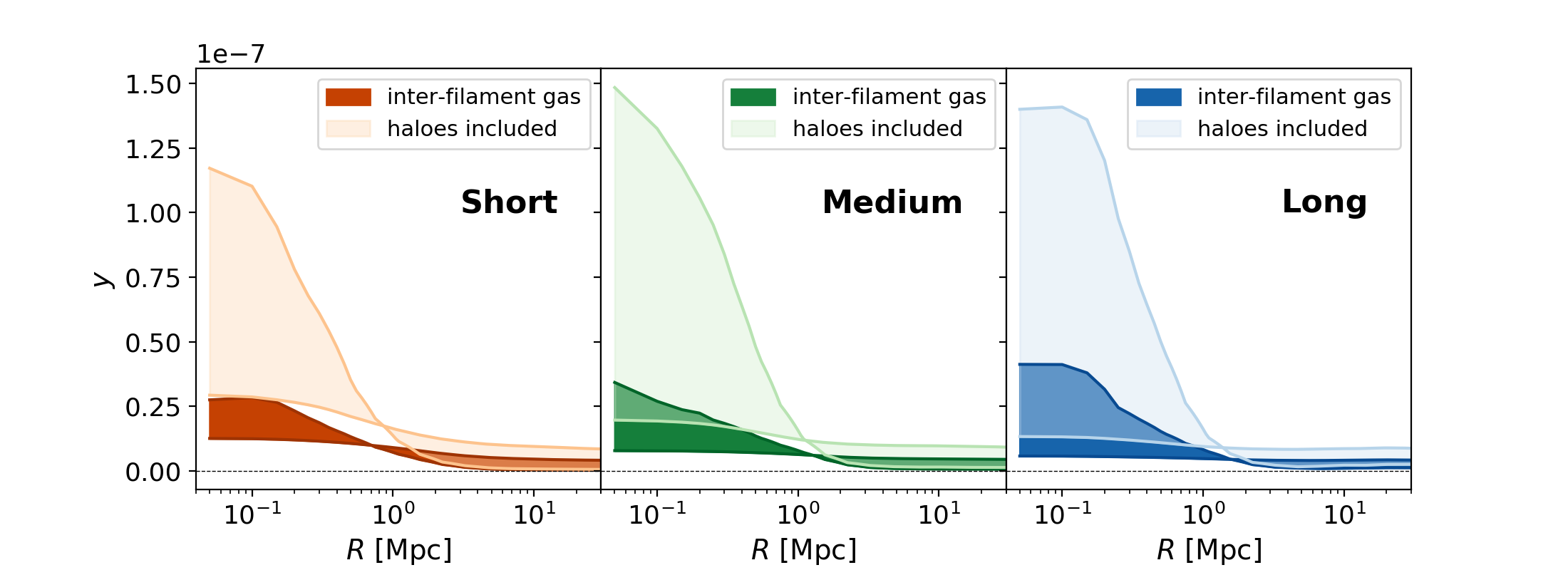}
   \caption{Compton-\textit{y} profiles of short ($L_f < 9$ Mpc), medium-length ($9 \leq L_f < 20$ Mpc), and long filaments ($L_f \geq 20$ Mpc) respectively in the left, middle and right panels. Dark colours show the SZ signal from inter-filament gas, i.e. gas that excludes the contribution of galactic haloes (massive $M_\mathrm{tot} > 10^{12} \, \mathrm{M}_\odot$ haloes hosting at least one galaxy of stellar mass $10^{9} \le \mathrm{M}_{*} \le 10^{12} \, \mathrm{M_{\odot}}$). Light colours correspond to the SZ signal in filaments where the contributions of galactic haloes have been included. The lower and upper limits are computed respectively with Eq.~\ref{Eq:y_min} and \ref{Eq:y_max}, using the mean filament lengths of each population: $\overline{L_f}= 5.7$ (short), 13.3 (medium) and $27.1$ Mpc (long filaments).}
    \label{Fig:y_pro_NoPhases}
    \end{figure*}

Gas pressure can be measured via the thermal Sunyaev-Zel'dovich (SZ) effect \citep{Zeldovich1969_SZ, Sunyaev1970_SZ, Sunyaev1972_SZ}. We therefore compute, from the mean pressure profiles of Fig.~\ref{Fig:Pres_pro_NoPhases}, the Compton \textit{y}-profiles of gas around filaments. 
We present two extreme cases of filament orientation with respect to the line of sight (l.o.s.): \textit{(i)} the filament is perpendicular to the l.o.s. (which gives a lower limit of SZ signal), and \textit{(ii)} the filament is parallel to the l.o.s. (upper limit of SZ signal).
In the following, we assume that filaments are far enough from the observer for all l.o.s. to be parallel.

For the perpendicular orientation, we compute the Compton-\textit{y} value at the distance $R$ from the spine of the filament ($R$ is the projected distance on the sky) using the following equation:
\begin{equation}\label{Eq:y_min}
    y(R) = \frac{\sigma_T}{m_e \, c^2} \int_0^{+ \infty} 2 \,\, P \bigg(\sqrt{R^2 + l^2} \bigg) \, dl, 
\end{equation} 
where $\sigma_T$, $m_e$ and $c$ are the Thomson scattering cross-section, the electron mass, and the speed of light in vacuum, $P$ is the mean pressure profile of Fig.~\ref{Fig:Pres_pro_NoPhases}, and $l$ denotes the distance from the filament to the observer along the l.o.s..

In the second case, the filament is parallel to the l.o.s., and the Compton-\textit{y} parameter depends on filament length $L_f$, as shown by:
\begin{equation}\label{Eq:y_max}
    y(R) = \frac{\sigma_T}{m_e \, c^2} P (R) \, L_f.
\end{equation}
Given that we deal with mean profiles of each filament population, the filament length values that we use are
the average lengths: $\overline{L_f}= 5.7, 13.3$ and $27.1$ Mpc for short, medium-length and long filaments respectively.

The resulting Compton-\textit{y} profiles are shown in Fig.~\ref{Fig:y_pro_NoPhases} for short (left panel), medium-length (middle) and long filaments (right panel).
The curves corresponding to the lower and upper signal limits are computed respectively with Eq.~\ref{Eq:y_min} and \ref{Eq:y_max}, and so the expected SZ signal from the various orientations should be bracketed between these curves, in the colour-filled regions.
Dark colours show the results from inter-filament gas excluding galactic haloes, while light colours correspond to the results including the contributions of these structures (as in Sect.~\ref{Sect:ContributionGalaxies}).

At cores of filaments, we see that the values of the SZ signal for the perpendicular orientation are the largest in short filaments. This is due to the high pressures and the wider shape of the pressure profile of this population with respect to the other, medium-length and long filaments. Concerning the parallel orientation, in the case where the contribution from galactic haloes is removed (dark colours), the maximum SZ signal of the inter-filament medium is associated with long filaments, with $y = 4.1 \times 10^{-8}$ at their core (see dark-blue upper limit), as expected from a signal that is proportional to filament length (Eq.~\ref{Eq:y_max}).
Interestingly, we note that the upper values in cores of short and medium-length filaments, $y \sim 2-3 \times 10^{-8}$, are close to that of the long-filament population, and this is a result of the larger pressure values of these populations (see Fig.~\ref{Fig:Pres_pro_NoPhases}), that compensate the smaller filament lengths.

We note that the inclusion of gas in galactic haloes (light colours) increases significantly, by a factor three, the SZ signal, which is expected since these structures are hotter and denser and might be strong SZ sources by themselves. Indeed, when galactic haloes are taken into account, the resulting SZ signal at filament cores is around $y \sim 1.2 \times 10^{-7}$ for the short population, $y \sim 1.5 \times 10^{-7}$ for medium-length, and $y \sim 1.4 \times 10^{-7}$ in long filaments. Maximum values are proportional to both pressure and filament length, so the similar \textit{y} values in medium-length and long filaments are due to the higher pressures in the former, and to the longer lengths of the latter.\\

    \begin{figure}
   \centering
   \includegraphics[width=0.5\textwidth]{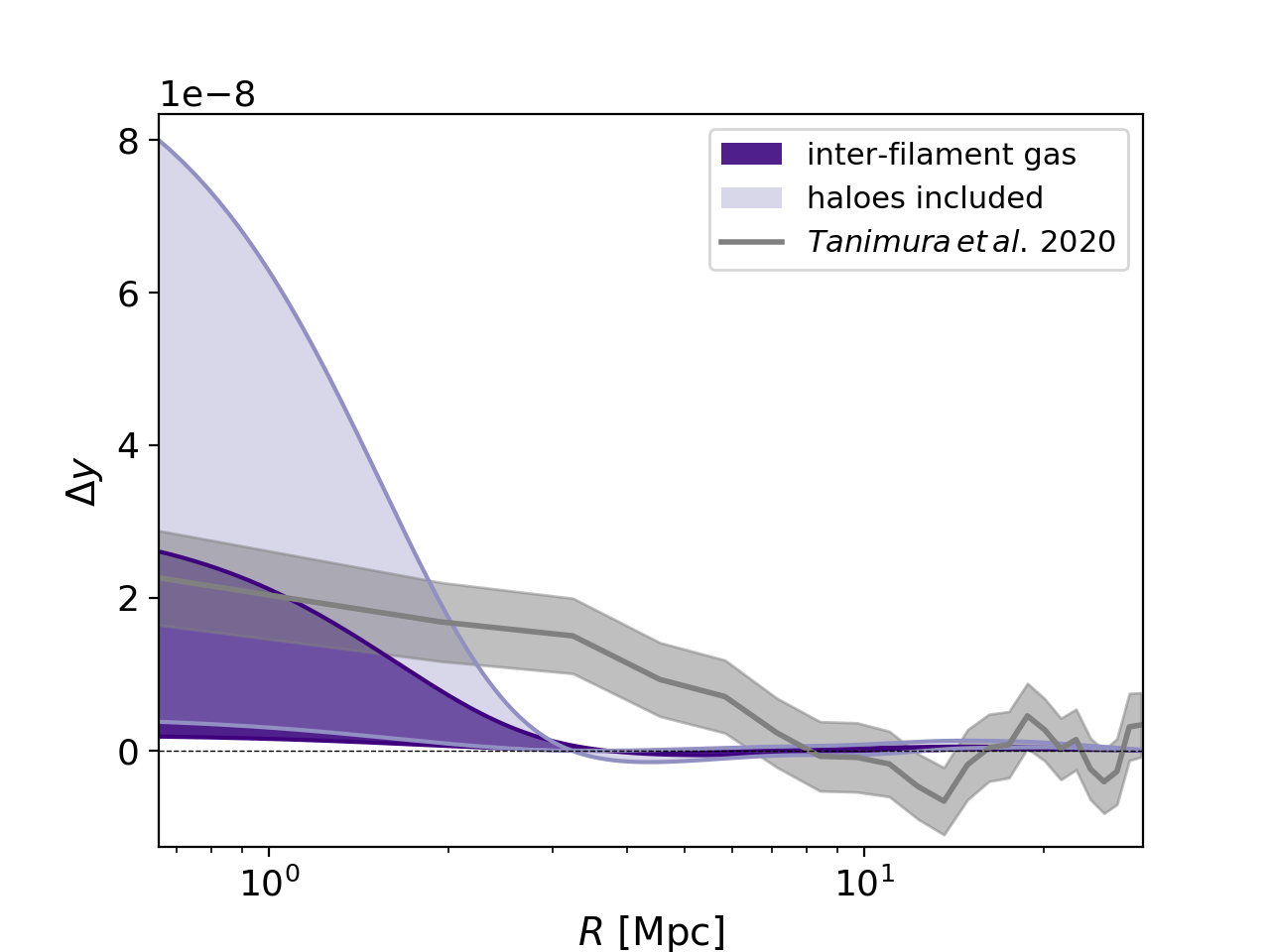}
   \caption{Mean $\Delta y$ profile of long filaments ($L_f \geq 20$ Mpc), convolved with the \textit{Planck} beam at $z=0.4$. Dark purple colours show the results of gas in the inter-filament medium (i.e. excluding the contribution of galactic haloes), while light purple colours present our findings when galactic haloes are included. The lower and upper boundaries (solid purple lines) are computed respectively using Eq.~\ref{Eq:y_min} and \ref{Eq:y_max}. We compare our results from simulations with observations from \cite{Tanimura2020_byopic}, which are shown in gray (line and $1 \sigma$ errors).}
    \label{Fig:y_pro_obs}
    \end{figure}

The SZ signal between pairs of clusters has been measured in \textit{Planck} data, but the comparison of these results to our findings from simulations faces intrinsic limitations (e.g. instrumental noise, density of sources, foregrounds, etc.).
It is however remarkable that our ranges of expected \textit{y} values from filaments including haloes are of the same order of magnitude as SZ measurements around luminous red galaxies of mass $M_* > 10^{11.3}$ $\mathrm{M}_\odot$ \citep{Tanimura2019}. We also compare our results with the measured SZ signal from the bridge between the clusters A399-A401 of \cite{Bonjean2018}. 
The signal in this filament was found to be remarkably high, with a value of the Compton-\textit{y} parameter of $\sim 10^{-5}$. This is about two orders of magnitude higher than our predicted ranges in filaments (including the contribution of haloes), which is not surprising from this exceptional bridge.

We finally perform a more complete comparison with the recent measurements at $4.4  \sigma$ significance of SZ signal from hot ionised gas around large-scale filaments presented in \citet{Tanimura2020_byopic}. 
In that paper, the authors built an average Compton-\textit{y} profile by stacking the \textit{Planck} 2015 \citep{PlanckColab2016_XXII} \textit{y} map, after removing the contribution of known groups and clusters, around filaments detected in the Sloan Digital Sky Survey (SDSS) by \cite{Malavasi2020_sdss}. In order to reproduce the observational constraints of \cite{Tanimura2020_byopic}, we use the mean \textit{y} profile we obtained for the long filament population. This is the closest possible to the sample of long ($30-100$ Mpc) filaments used in \cite{Tanimura2020_byopic}. We indeed note that, within the DisPerSE framework, filament lengths strongly depend on the density of tracers. The filament catalogue in \cite{Malavasi2020_sdss}, from SDSS, and in our work are derived with $4 \times 10^{-4}$ and $10^{-2}$ galaxy/Mpc$^3$ respectively. As a consequence, the filament length distribution of the SDSS catalogue misses a lot of short filaments and thus peaks at $20-30$ Mpc. 
We rescale our mean \textit{y} profile of long filaments to the average redshift of $z=0.4$ of the observed filament sample. 
We also convolve the \textit{y} profile by a Gaussian beam of 10 arcmin at redshift $z=0.4$, corresponding to resolution of the SZ \textit{Planck} map analysed in \citet{Tanimura2020_byopic}.
Finally, following their approach, we compute the excess of SZ signal $\Delta y$ with respect to the background by dividing the \textit{y} profile derived from simulations by its background value, defined as the mean value at distances $r>25$ Mpc from the spine. 

The resulting $\Delta y$ profiles are displayed in Fig.~\ref{Fig:y_pro_obs} in light and dark purple colours, while the measurement from \cite{Tanimura2020_byopic} is shown in gray (lines and shaded regions of $1 \sigma$ errors).
For inter-filament gas (dark purple), we find that the maximum SZ signal derived from the IllustrisTNG simulation is expected to be $\Delta y = 2.6 \times 10^{-8}$, compatible within the $1 \sigma$ with the \cite{Tanimura2020_byopic} profile. Given that our values correspond to upper bounds and that the observed filaments have a distribution of orientations, we deduce that a small contribution from the galactic haloes is expected in the measured signal, as it is suggested by the predicted signal of filaments including haloes (light purple curve) peaking at $\Delta y = 8.0 \times 10^{-8}$.
Moreover, we note that the differences in the radial extent of the profiles could be explained by the reduced number of tracers in the SDSS catalogue (as mentioned above). The low galaxy density indeed increases the uncertainties on the position of the filament spines, thus broadening the cores of the DisPerSE filaments \citep[as shown in][]{GalarragaEspinosa2020}. However at this stage, we are not able to distinguish this effect from possible thicker filament cores that would be expected at higher redshifts.

\section{\label{Sect:Conclusions}Conclusions}

Using the TNG300-1 simulation at redshift $z=0$, we have analysed the gas distribution and properties around the filamentary structures of the cosmic web. We distinguished five different gas phases, according to their density and temperature in the phase diagram, and we have analysed their distribution around three different filament populations characterised by their length. The main results of this work are listed in the following.

\begin{enumerate}[$\bullet$]
    \item We have found that cores of filaments ($r \leq 1$ Mpc) contain essentially WHIM gas, but the hotter and denser phases of Hot gas, and also WCGM, are also significantly present (see Fig.~\ref{Fig:gas_phase_fraction} and Table~\ref{Table:fractions_core}).
    However, cores of short and long filaments are not made of exactly the same gas. Short filaments contain hotter gas than the long population, and long filaments possess a contribution of cold and diffuse IGM gas at their cores, that is absent in the short filament population (see phase-space of Fig.~\ref{Fig:PS_3cols}). These differences can be explained by the different large scale environments (that are denser and less-dense respectively for short and long filaments).

    \item We have found that at $r \sim 1$ Mpc from the spine of filaments, WHIM gas completely dominates ($>80 \%$) the entire baryon budget, and the other phases are negligible. The different gas content at cores and on the outskirts of filaments suggests that, apart from gravitational interactions, additional baryonic processes (whose efficiencies depend on the type of filament) affect the gas located at the cores, making it significantly different from the gas located further away from the spine. These additional baryonic processes might be due to the enhanced presence of haloes in these regions \citep{GalarragaEspinosa2020}, that can accrete gas to their gravitational potential well, but also expel it by other feedback effects.
    
    \item We have estimated that the average temperature and pressure at cores of filaments ($r \leq 1$ Mpc) are $T = 4 - 13 \times 10^{5}$ K, and $P = 4 - 12 \times 10^{-7}$ $\mathrm{keV.cm}^{-3}$, depending on the cosmic environment (see Table~\ref{Table:TPtable}). Short filaments (living in denser regions) have temperature and pressure values that are three times those of the long population (which traces less-dense environments). 
    All filaments present isothermal cores up to distances of $r_\mathrm{core} = 1.5$ Mpc (see Fig.~\ref{Fig:Temp_pro_NoPhases}), which is in agreement with previous studies \citep{KlarMucket2012, GhellerVazza2019_surveyTandNTprops_fils}.
    Finally, we observed that pressures in filament cores are $\sim 1000$ times lower than those in cores of clusters.
    
    \item We have estimated the SZ signal associated with gas in filaments, and we have found it in the range $y = 0.5 - 4.1 \times 10^{-8}$, depending on the type of filament (see Fig.~\ref{Fig:y_pro_NoPhases}). When gas from galactic haloes is included, the expected SZ signal increases significantly. In this case, the Compton-\textit{y} parameter is in the range $y = 0.1 - 1.5 \times 10^{-7}$.
    We have compared with the recent observations of gas in filaments \citep{Bonjean2018, Tanimura2019, Tanimura2020_byopic} and we have found compatible SZ ranges with the expected signal of gas (Fig.~\ref{Fig:y_pro_obs}). 
    
    \item We have shown that diffuse phases (Diffuse IGM and WHIM) are extremely sensitive to the large scale cosmic environment, traced by short and long filaments (see their corresponding profiles in Fig.~\ref{Fig:Temp_pro} and \ref{Fig:Pres_pro}).
    On the outskirts of filaments ($r > 1$ Mpc) where these phases are the most present, there are statistically less haloes than in filament cores \citep{GalarragaEspinosa2020}, and so gas in these rather empty regions might be mainly shaped by the gravitational pull towards the denser regions of the cosmic web. Gravity is thus probably the main responsible for the rise of temperature of the accreted Diffuse IGM (see Fig.~\ref{Fig:Temp_pro}), that is therefore converted into WHIM gas (Fig.~\ref{Fig:gas_phase_fraction}) when approaching the spine of the filament. On the contrary, we have found that the hotter and denser phases (WCGM, Halo and Hot gas) are almost insensitive to the global environment of short and long filaments. Indeed, given that these phases are associated with haloes (Fig.~\ref{Fig:PS_3cols}), their properties might be rather shaped by smaller scales (e.g. accretion onto haloes, halo mergers) and by  galactic physics (e.g. feedback effects).
    Moreover, haloes are (biased) tracers of the density field, and they are thus widely present in cores of filaments \citep{GalarragaEspinosa2020}. Accordingly, we have found that all the gas phases at filament cores ($r \leq 1$ Mpc) are significantly impacted by the higher number density of these collapsed structures (see Fig.~\ref{Fig:Temp_pro} and \ref{Fig:Pres_pro}).
    
\end{enumerate}

The physics of gas in the cosmic web is complex, and since baryonic matter is the only component that can be directly observed, it is essential to characterise its distribution and properties. In this sense, this work allowed us to build a clearer picture of the physical state of gas at any distance from the spine of filaments, and to identify more precisely the scales at which the different processes that shape gas in the cosmic web enter the game. 
Of course, further studies of the dynamics of gas and its evolution across redshifts are necessary in order to have a comprehensive understanding of the properties of gas in filaments, including dedicated mapping of the balance of power between the processes at play as function of time and environment in the cosmic web.




\begin{acknowledgements}
The authors thank the anonymous referee for helpful comments.
This research has been supported by the funding for the ByoPiC project from the European Research Council (ERC) under the European Union’s Horizon 2020 research and innovation program grant
agreement ERC-2015-AdG 695561. (ByoPiC, https://byopic.eu). The authors acknowledge the very useful comments and discussions with all the members of the ByoPiC team (https://byopic.eu/team/), and thank Marius Cautun for providing constructive feedback. DGE is especially grateful to  Victor Bonjean and C\'eline Gouin for stimulating discussions. We thank the IllustrisTNG team for making their data publicly available.
\end{acknowledgements}


\begin{appendix}

\section{\label{Appendix:Resolution} Impact of the resolution of simulation}

We study the effects of resolution on our results by performing the same analysis on the TNG300-2 simulation box. This box is the medium resolution run of the TNG300 series, so it has the same characteristics as the TNG300-1 simulation, our baseline, except for an eight times lower mass resolution ($m_{\mathrm{DM}} = 3.2 \times 10^{8} \mathrm{M_{\odot}}/h$), a target baryon mass of $m_{\mathrm{baryon}} = 5.9 \times 10^{7} \mathrm{M_{\odot}}/h$, and a reduced number of gas cells $N_\mathrm{gas} =1250^3$.

    \begin{figure*}
   \centering
   \includegraphics[width=1\textwidth]{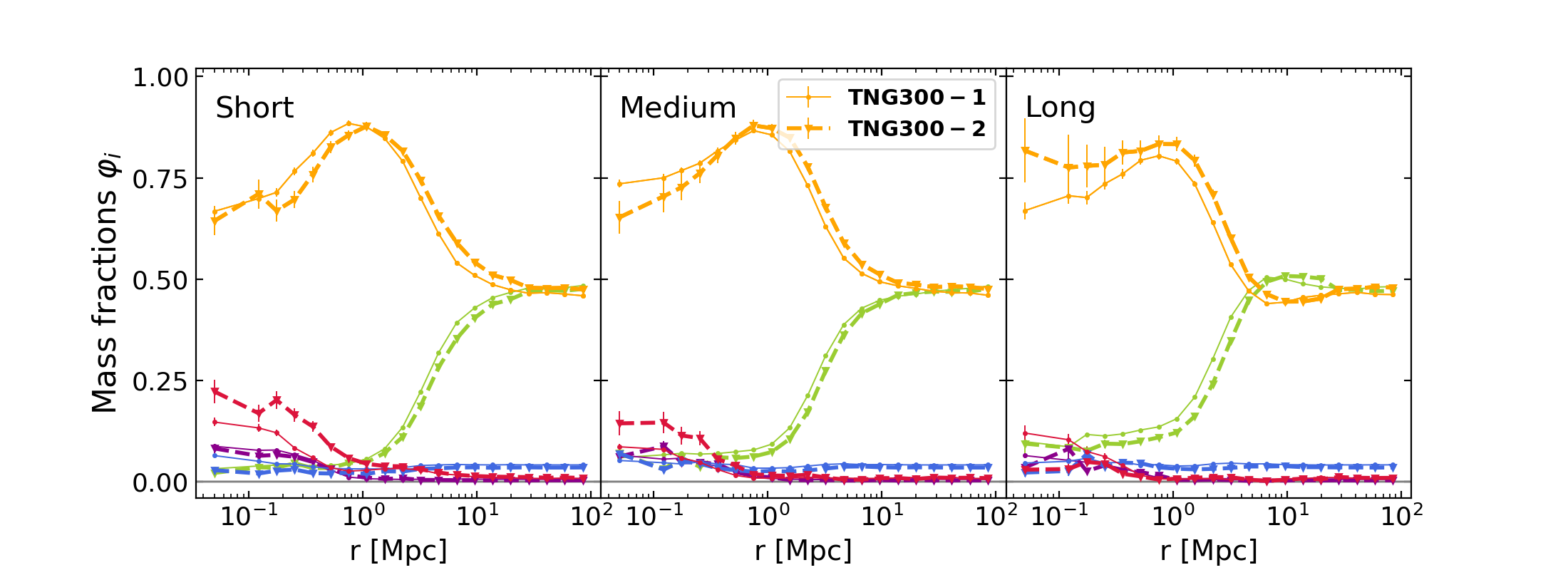}
   \caption{Gas mass fraction profiles $\varphi_i$ (see Eq.~\ref{Eq:phi_i}) of the five different gas phases in the TNG300-2 simulation (thick dashed lines), and in the reference TNG300-1 box (thin solid lines, same as Fig.~\ref{Fig:gas_phase_fraction}). The colour codes are these of Fig.~\ref{Fig:gas_phase_fraction}, so that Diffuse IGM, WHIM, WCGM, Halo and Hot gas correspond respectively to the green, yellow, purple, blue and red curves.}
    \label{Fig:gas_phase_fraction_RESOLUTION}
    \end{figure*}
    
        \begin{figure}
   \centering
   \includegraphics[width=0.5\textwidth]{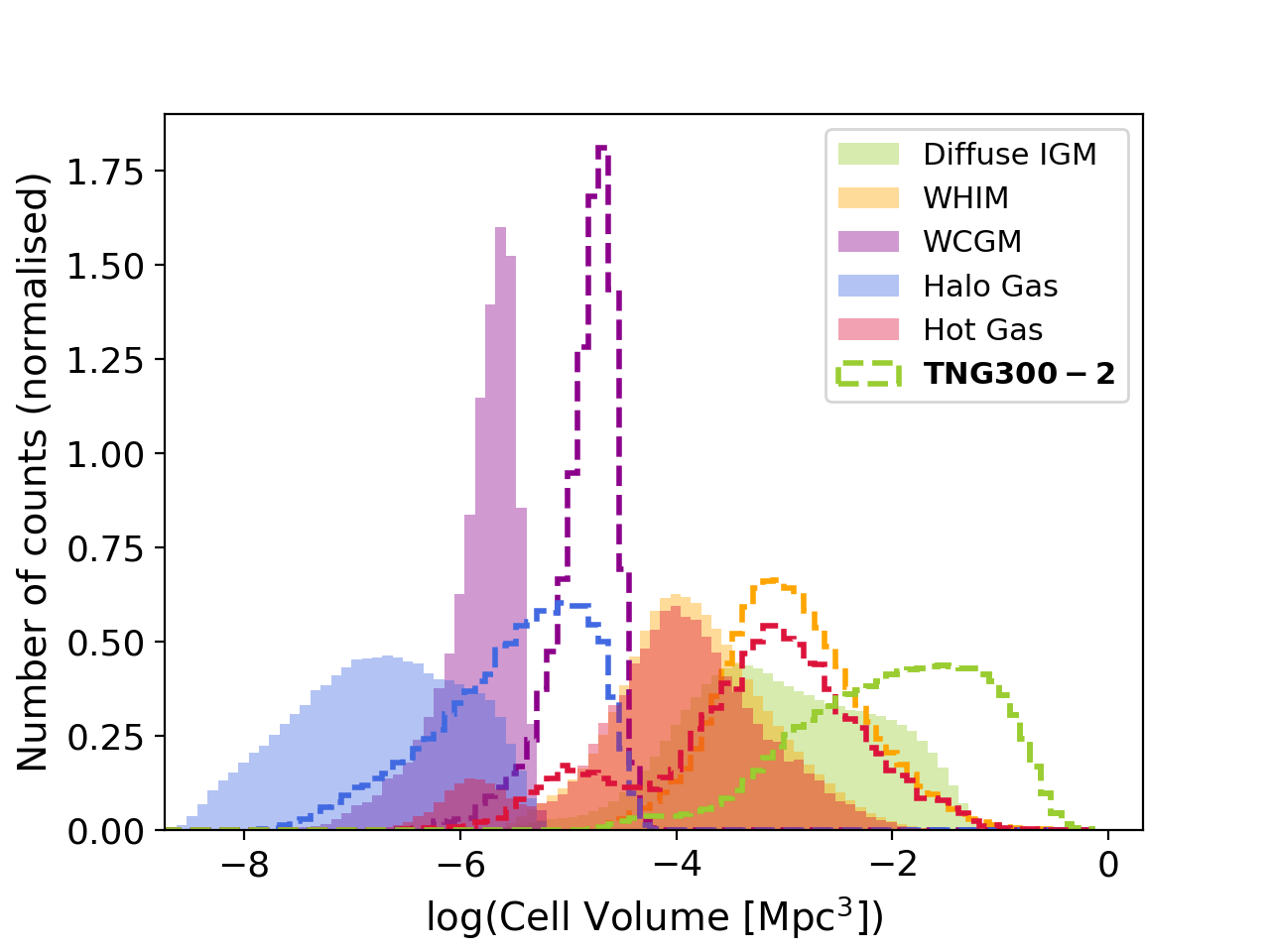}
   \caption{Distribution of the volumes of the gas cells in the TNG300-1 box (colour-filled histograms) and in the TNG300-2 simulation (dashed lines). The different colours correspond to the different gas phases considered in this work.}
    \label{Fig:VolCell_2SIMS}
    \end{figure}
    
Figure \ref{Fig:gas_phase_fraction_RESOLUTION} presents our findings concerning the mass fractions profiles $\varphi_i$ of the different gas phases in these two simulations. The results of the reference TNG300-1 simulation (previously shown in Fig.~\ref{Fig:gas_phase_fraction}) are presented in thin solid lines, and we compare to them the $\varphi_i$ profiles from the TNG300-2 box (in thick dashed lines).
The profiles of TNG300-1 and TNG300-2 are quite similar, except for a very slight difference in the $\varphi_i$ amplitudes, that is the most noticeable at the core of filaments ($r \leq 1$ Mpc). This difference is explained by the coarser Voronoi grid of the TNG300-2 box. Indeed, Fig.~\ref{Fig:VolCell_2SIMS} clearly shows that the distribution of the cell volumes in the TNG300-2 simulation is considerably shifted towards the larger values with respect to the TNG300-1 distributions, and this is the case for all the gas phases. As a consequence, the coarser grid of the TNG300-2 box leads to a slightly different distribution of the gas cells in the phase-space plane, meaning that, unsurprisingly, the classification of the gas cells into the different phases is less precise in this simulation.

\begin{figure}
   \centering
   \includegraphics[width=0.5\textwidth]{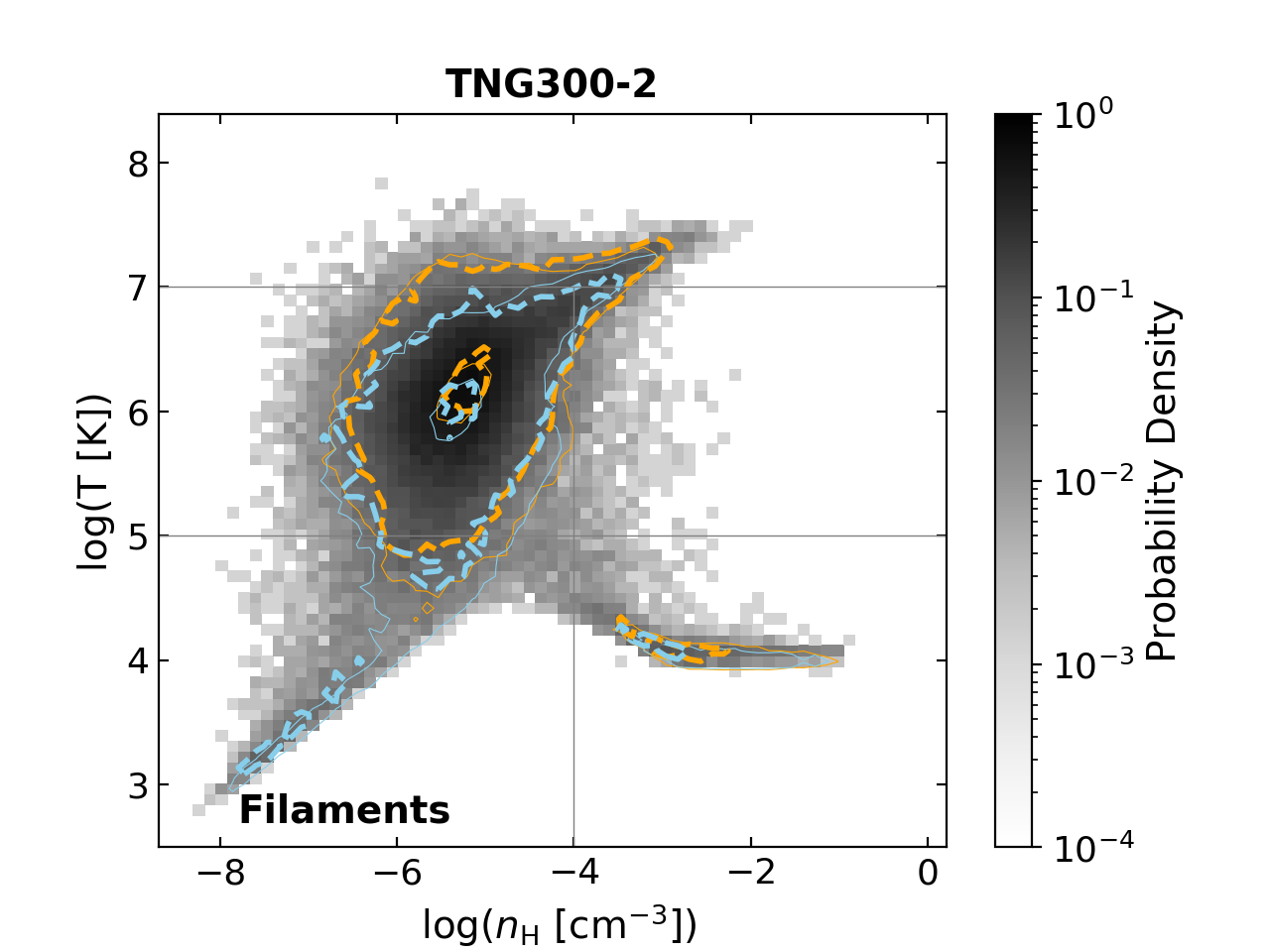}
   \caption{Phase-space diagram of cells at cores of filaments ($r \leq 1$ Mpc) in the TNG300-2 simulation. The specific contours of short and long filaments are shown by the thick dashed lines, respectively in orange and blue. The results from TNG300-1 (presented in Fig.~\ref{Fig:PS_3cols}) are over-plotted in thin solid lines.}
    \label{Fig:PS_fils_2SIMS}
    \end{figure}

For quantitative purposes, we show in Table~\ref{Table:fractions_2SIMS} the total fractions of the five different gas phases in our two simulations (we consider here the full simulation boxes, excluding nodes and haloes). This table shows that, indeed, the TNG300-2 box possesses slightly different fractions of cells with respect to the reference TNG300-1, with larger fractions in the hotter phases (e.g. Hot gas and WHIM), and reduced ones in the cooler phases (e.g. Diffuse IGM, Halo gas). As expected, this is reflected in the phase-space of gas at cores of filaments ($r \leq 1$ Mpc) that we present in Fig.~\ref{Fig:PS_fils_2SIMS}. Here, the contours of the TNG300-2 simulation corresponding to the gas content of short and long filaments are slightly shifted in comparison to these of TNG300-1. Nevertheless, all the differences described above remain very tiny.\\

\begin{table}
\caption{Fractions of cells in the five different gas phases considered in this work, for the full TNG300-1 and TNG300-2 simulations. These numbers exclude the gas cells lying within spheres of radius $3 \times R_{200}$ centered at the positions of CPmax (i.e. the topological nodes) and of galactic haloes, and thus focus only on the inter-filament gas of the simulation.}
\label{Table:fractions_2SIMS}     
\centering  
\begin{tabular}{ c | c c }
 \hline\hline  
     & TNG300-1 & TNG300-2 \\ \hline
    Diffuse IGM & $48.0 \% $ & $47.2 \%$ \\
    WHIM & $46.8\%$ & $48.3 \%$ \\
    WCGM & $0.6 \%$ & $0.4 \%$ \\
    Halo Gas & $3.7 \%$ & $3.0 \%$ \\
    Hot Gas & $0.9 \%$ & $1.1 \%$ \\
  \hline
 \end{tabular}
\end{table}

    \begin{figure*}
   \centering
   \includegraphics[width=0.5\textwidth]{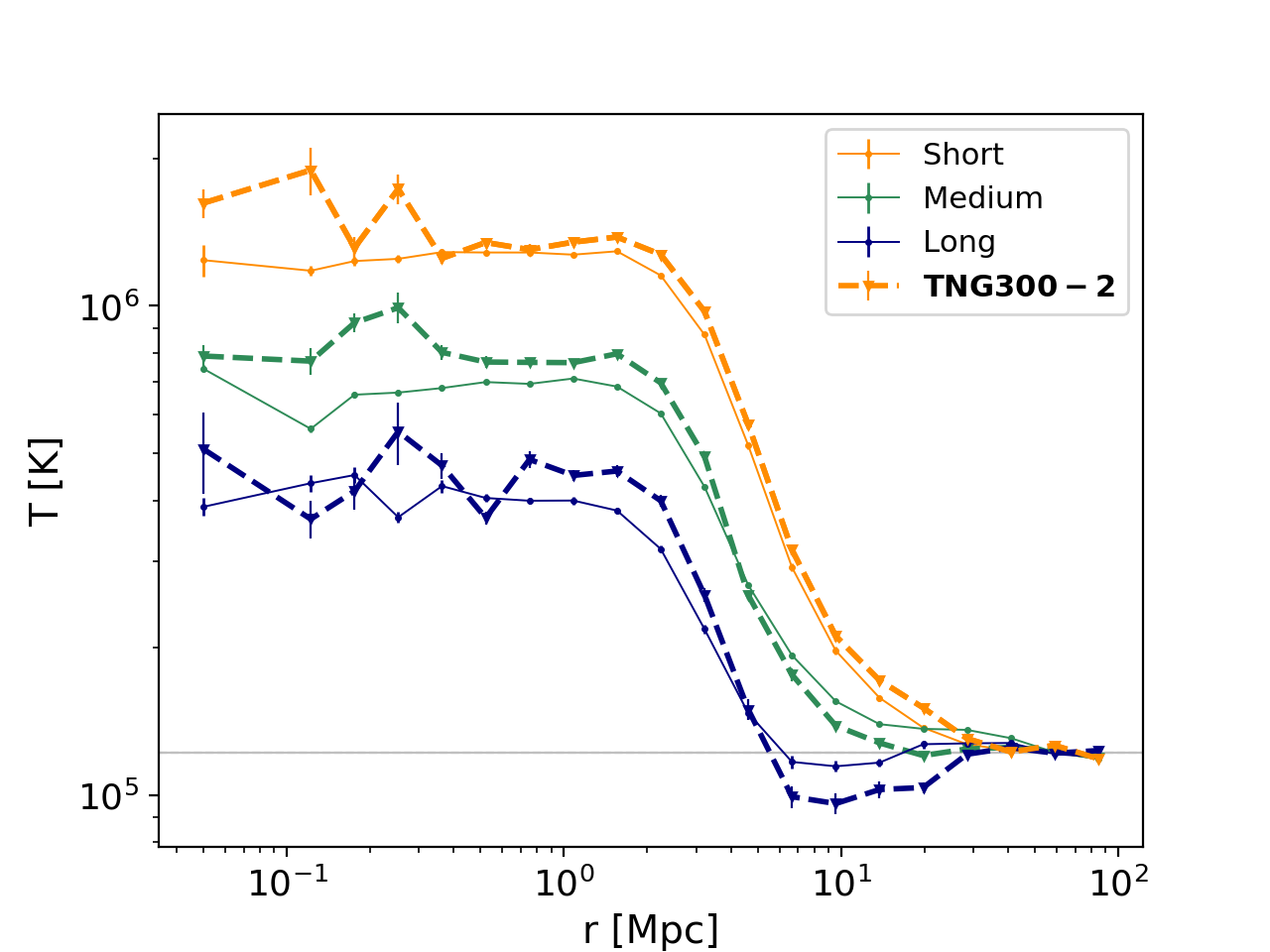}\includegraphics[width=0.5\textwidth]{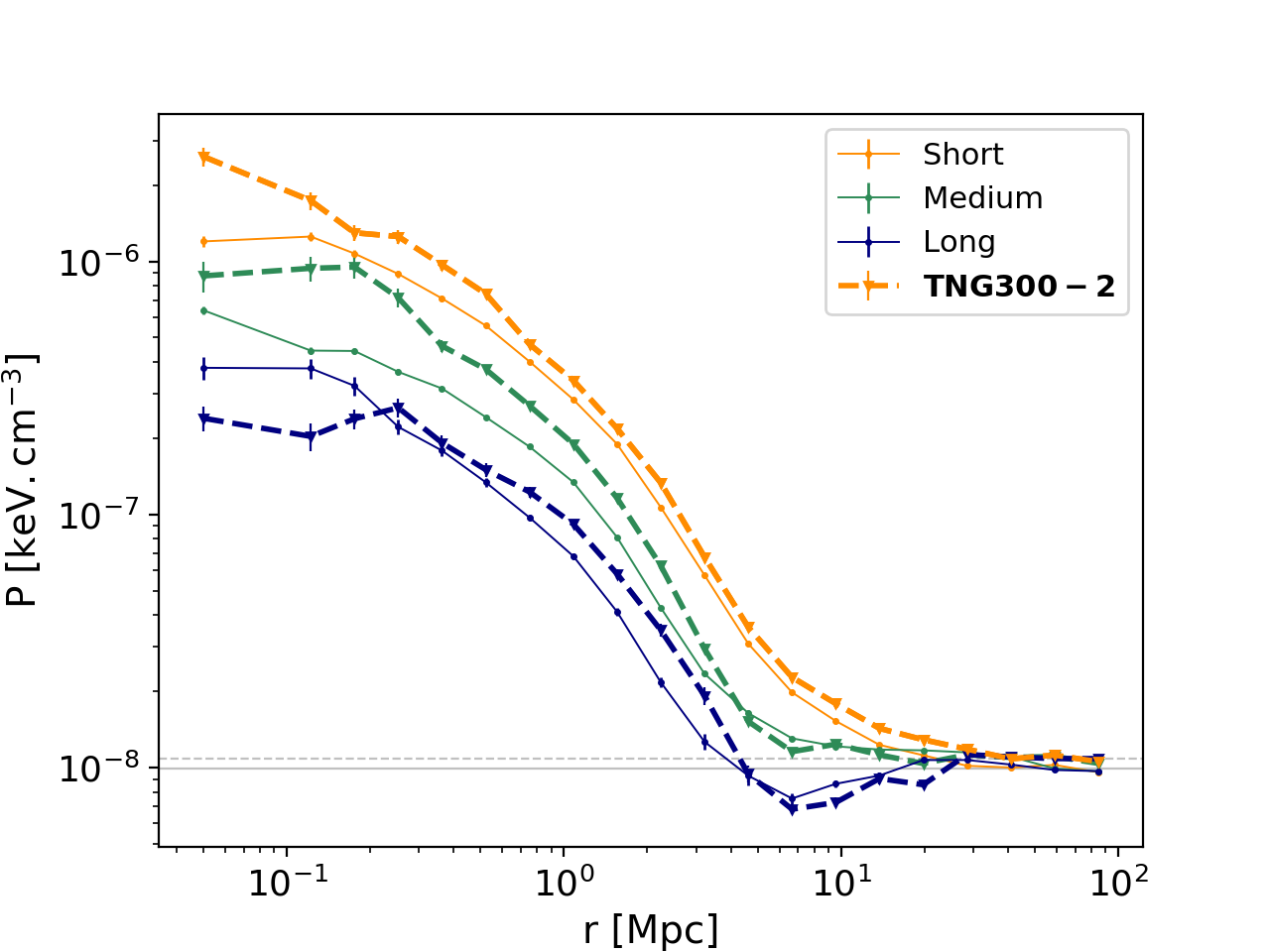}
   \caption{Temperature and pressure profiles of short, medium-length, and long filaments in the TNG300-1 and TNG300-2 simulations, respectively in thin solid, and in thick dashed lines. We note that the results from TNG300-1 are the same as presented in Fig.~\ref{Fig:Temp_pro_NoPhases} and \ref{Fig:Pres_pro_NoPhases}.}
    \label{Fig:profilesTP_RESOLUTION}
    \end{figure*}
    
We now investigate the effects of resolution in the filament temperature and pressure profiles. We compare in Fig.~\ref{Fig:profilesTP_RESOLUTION} the profiles of the TNG300-2 filaments to those of the reference TNG300-1 skeleton.
Again, we observe only minor differences, that consist mainly in slightly hotter profiles and higher pressure values in the TNG300-2 curves. This only reflects the previously discussed features, that arise from the coarser Voronoi grid of the TNG300-2 simulation.\\

It is important to note that the characteristic radial scales of gas around filaments that we have discussed in the main text are found to be resolution independent, since all the results from the TNG300-1 and TNG300-2 simulations exhibit the same radial features, i.e. maximum of WHIM at $r \sim 1$ Mpc and isothermal cores up to $r \sim 1.5$ Mpc (see Sect.~\ref{SubSect:gas_fraction_profiles} and \ref{SubSect:T_profiles}), despite the TNG300-2 box having a resolution that is eight times lower than the reference. Therefore, this study shows that the resolution of the simulation has only minor effects in the gas distribution and properties around cosmic filaments, as expected, since we focus here on the largest scales of the Universe. The main conclusions presented in this paper are therefore free from resolution effects.

\end{appendix}

\bibliography{main} 

\end{document}